\documentclass[reprint,twocolumn,superscriptaddress,showkeys,
nofootinbib,notitlepage,amsmath,amssymb,floatfix,aps,prd]{revtex4-2}

\usepackage{latexsym,amsmath,amssymb,lmodern,url}
\usepackage{natbib}
\usepackage{color}
\usepackage{multirow}
\usepackage{bm}
\usepackage{array}
\usepackage[normalem]{ulem}
\usepackage{cleveref}
\usepackage{xcolor}

\usepackage{tikz}
\usetikzlibrary{datavisualization}
\usetikzlibrary{datavisualization.formats.functions}
\usetikzlibrary{plotmarks}

\usepackage{mathtools}

\def\de{\delta^{\vphantom{1}}}
\def\bde{{\bar\delta}}

\def\h3{{\displaystyle{\frac 3 2}}}

\begin{document}
\title{Fine Structure and Decays of Hidden-Strangeness Tetraquarks in the Dynamical Diquark Model}
\author{Shahriyar Jafarzade}
\email{shah.jzade@pitt.edu}
\affiliation{University of Pittsburgh, Pittsburgh, PA 15260, USA}
\affiliation{Department of Physics, Arizona State University, Tempe,
AZ 85287, USA}
\author{Richard F. Lebed}
\email{Richard.Lebed@asu.edu}
\affiliation{Department of Physics, Arizona State University, Tempe,
AZ 85287, USA}
\date{March, 2026}

\begin{abstract}
We analyze the fine structure and ``fall-apart'' decay patterns of hidden-strangeness tetraquarks within the dynamical diquark model.  Several well-established negative-parity resonances listed by the Particle Data Group (PDG) [$\phi(2170)$, $\eta(2225)$, $\eta(2370)$] are examined as potential $q\bar q s\bar s$ tetraquark candidates using a Hamiltonian that incorporates (iso)spin-dependent, spin–orbit, and tensor interactions.  We further show that the isovector resonances $\rho(2150)$ and $\rho_3(2250)$ observed by BESIII in $\psi(2S)\rightarrow K^{+}K^{-}\eta$ are compatible with a tetraquark assignment.  Predictions for 28 states, including those with exotic quantum numbers, are also presented.  Comparison of the model spectrum with the PDG's unverified ``Further States’’ highlights additional promising candidates worth future experimental investigation.  The predicted fall-apart decay channels are easily reconstructed by experiment, and may stimulate ongoing searches for hidden-strangeness tetraquarks at GlueX and BESIII\@.   While current data does not yet provide an unambiguous interpretation for any state in this range, the observation of the fall-apart modes, in addition to the distinct multiplet structure for tetraquark states from that of hybrids or glueballs, provide the most incisive tests for discriminating state content. 
\end{abstract}

\maketitle

\section{Introduction}

The spectroscopy of hadrons provides a crucial window into the dynamics of 
quantum chromodynamics (QCD) at low energies~\cite{Gross:2022hyw}.  While the majority of observed resonances are successfully described within the conventional quark model as either quark-antiquark mesons or as three-quark baryons, the spectrum of QCD is not restricted to these simple configurations.  The past two decades have shown that multiquark states and gluonic excitations represent an integral part of 
hadronic physics, most clearly demonstrated in the heavy-quark sector through 
the discovery of numerous unconventional charmoniumlike and bottomoniumlike 
states.
These developments motivate a broader and more 
systematic examination of exotic spectroscopy across all flavor sectors.

The strange-quark sector, in particular, presents several long-standing 
puzzles.  Several resonances observed above $2~\text{GeV}$ exhibit properties that 
cannot be easily reconciled with a conventional $q\bar{q}$ interpretation.  This mass region is especially challenging to analyze because the masses predicted for the lightest glueball and hybrid multiplets also lie nearby.  Consequently, the observed negative-parity states may embody significant mixing between conventional mesons, glueballs, hybrids, and tetraquarks, making their identification highly nontrivial.

Existing theoretical studies underscore the significance of this mass region for understanding exotic-hadron QCD dynamics.  A lattice calculation of the light-meson spectrum~\cite{Dudek:2013yja} does not resolve a 
sufficient number of hybrid states to reliably determine the spin-averaged mass of the expected $H_1$ supermultiplet, whose members have quantum numbers $1^{--}$ and $(0,1,2)^{-+}$.  Complementary analyses based upon a constituent-gluon model~\cite{Swanson:2023zlm} place the spin-averaged $H_1$ mass near $2.2~\text{GeV}$, and comparison with lattice results suggests an isovector/vector hybrid mass of about $2.1~\text{GeV}$, while the corresponding isoscalar/vector hybrid is expected to lie in the range 
$2.10$-$2.25~\text{GeV}$.  The hadronic decay patterns of 
these hybrid candidates have been studied in detail in Ref.~\cite{Farina:2023oqk}.
 
At the same time, the dynamical diquark model, originally developed in Refs.~\cite{Brodsky:2014xia,Lebed:2015tna,Lebed:2017min} to 
describe exotic multiquark hadrons in the heavy-quark sector, and found to successfully explain a wide range of tetraquark and pentaquark phenomena~\cite{Brodsky:2014xia,Lebed:2017min,Giron:2019bcs,Giron:2019cfc,Giron:2021fnl,Giron:2020qpb,Giron:2020wpx,Gens:2021wyf,Giron:2021sla,Lebed:2015tna,Lebed:2022vks,Mutuk:2022nkw,Lebed:2023kbm,Mutuk:2024vzv,Lebed:2024rsi,Lebed:2024zrp,Lebed:2025xbz}, predicts a hidden-strangeness ($s\bar s q\bar q$) negative-parity tetraquark multiplet $\Sigma_g(1P)$ with spin-averaged mass near $2.3~\text{GeV}$~\cite{Jafarzade:2025qvx}.  The focus upon strangeness reflects the role of the $s$ quark as intermediate between the heavy-quark ($c,b$) sector in which states in the mass spectrum are cleanly separated, and the light-quark ($u,d$) sector in which they are not.  The proximity of hybrid and tetraquark mass predictions in this region highlights a central question: 
\emph{Which, if any, of the experimentally observed negative-parity states can be conclusively identified as hidden-strangeness tetraquarks?}

\begin{figure}[h]
    \centering
    \includegraphics[scale=0.65]{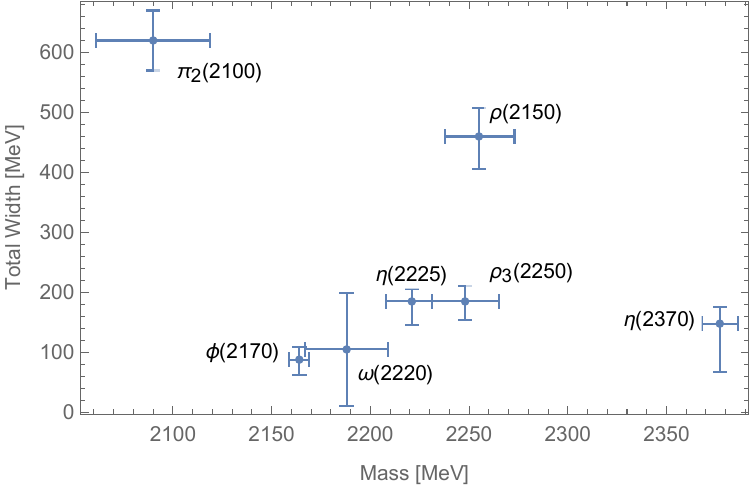}
    \caption{List of negative-parity resonances tabulated by the PDG~\cite{ParticleDataGroup:2024cfk} with mass above 2.1~GeV (and below $m_{\eta_c} = 2.984$~GeV).  Masses and total decay widths of $\rho(2150)$ and $\rho_3(2250)$ are taken from BESIII measurements in the channel $\psi(2S)\rightarrow K^{+}K^{-}\eta$~\cite{BESIII:2019dme}.}
    \label{fig:resonances}
\end{figure}

Several resonances listed by the Particle Data Group (PDG)~\cite{ParticleDataGroup:2024cfk} in the region above
$2.1 ~\text{GeV}$ exhibit negative parity and uncertain internal 
structure, including $\phi(2170)$, $\omega(2220)$, $\eta(2225)$, 
$\eta(2370)$, $\rho(2150)$, $\rho_3(2250)$, and $\pi_2(2100)$ 
(see Fig.~\ref{fig:resonances}).  Many of these states decay into channels 
containing strange mesons, suggesting a sizable hidden-strangeness component~\cite{Jafarzade:2025txm}.  Di-meson effects within the {\it diabatic\/} dynamical diquark model (an extension of the original model designed to incorporate di-meson coupled-channel effects~\cite{Lebed:2022vks,Lebed:2023kbm,Lebed:2024rsi,Lebed:2024zrp,Lebed:2025xbz}) are tested for some of these resonances in  Ref.~\cite{Jafarzade:2025qvx}.  In sharp contrast to the heavy-quark sector, these results find no hidden-strange tetraquark candidate whose structure is dominated by di-hadron structure.  
The masses and quantum numbers of these states therefore make them natural candidates for testing both the predictions of the dynamical diquark model beyond the heavy-quark sector, and for contrasting with the predictions of models for hybrid and excited conventional mesons.

In this work we examine whether the PDG-listed negative-parity resonances above 2.1~GeV and below the lowest hidden-charm state $\eta_c$ at 2.984~GeV can be understood as members of the hidden-strangeness $P$-wave tetraquark multiplet $\Sigma^+_g(1P)$ predicted by the dynamical diquark model.  We compute the fine-structure mass splittings, compare the resulting spectrum with experimental observations, and analyze the expected ``fall-apart'' decay patterns (2-body modes with no change in valence-quark content).  From this analysis, we assess the extent to which the dynamical-diquark framework provides a coherent description of the negative-parity hidden-strange meson sector, and sheds light upon the underlying multiquark structure of these resonances.

The paper is organized as follows.  Section~II introduces the $P$-wave states 
within the dynamical diquark model.  In Sec.~III we specify the fine-structure Hamiltonian and present our fit for the resulting mass spectrum.  Section~IV examines the fall-apart decay patterns for the hidden-strangeness tetraquarks, and Sec.~V summarizes our conclusions.

\section{$P$-wave Tetraquark States in the Dynamical Diquark Model}

Within the dynamical diquark model, a tetraquark is described as a bound state of a compact
diquark–antidiquark ($\delta\bar\delta$) pair, connected by a confining  gluonic flux tube.  $\de$ is taken to be a color antitriplet $(\mathbf{3} \otimes \mathbf{3} \rightarrow \mathbf{\bar 3)}$ and $\bde$ a color triplet $(\mathbf{\bar 3} \otimes \mathbf{\bar 3} \rightarrow \mathbf{ 3)}$, {\it i.e.}, the color channels that are attractive at short distance.  The $\de\bde$ pair then forms an overall color singlet.  This configuration admits an effective two-body description, in which the internal structure of each diquark is encoded in its spin quantum 
number, $s_{\de}$ or $s_{\bde}$.  The total spin $\mathbf{S} \equiv \mathbf{s}_{\de}+\mathbf{s}_{\bde}\,$ plays a central role in organizing the tetraquark spectrum.

The lowest-energy tetraquark states correspond to an $S$-wave configuration
($L=0$) between $\de$ and $\bde$.  These states all have positive parity (two quarks and two antiquarks), and together form the multiplet $\Sigma^+_g(1S)$ in the Born-Oppenheimer (BO) approximation~\cite{Lebed:2017min}.  The allowed total $J^{PC}$ quantum numbers follow entirely from including all possible $(s_\de, \, s_{\bde}, \, S)$ combinations, and respecting the symmetry of each state.  For completeness, the multiplet $\Sigma^+_g(1S)$, neglecting isospin, consists of the 6 states
\begin{align} \label{eq:Swave}
J^{PC}=0^{++}: \quad 
& X_{0} = \lvert 0_{\de}, 0_{\bde} \rangle_{0} \, , \;\;\;
  X_{0}^{\prime} = \lvert 1_{\de}, 1_{\bde} \rangle_{0} \, , 
\nonumber \\[2mm]
J^{PC}=1^{++}: \quad 
& X_{1} = \frac{1}{\sqrt{2}}
\Bigl(
\lvert 1_{\de}, 0_{\bde} \rangle_{1}
+
\lvert 0_{\de}, 1_{\bde} \rangle_{1}
\Bigr) ,
\nonumber \\[2mm]
J^{PC}=1^{+-}: \quad 
& \; Z \, = \; \frac{1}{\sqrt{2}}
\Bigl(
\lvert 1_{\de}, 0_{\bde} \rangle_{1}
-
\lvert 0_{\de}, 1_{\bde} \rangle_{1}
\Bigr) , 
\nonumber \\[2mm]
& \, Z^{\prime} = \lvert 1_{\de}, 1_{\bde} \rangle_{1} \, ,
\nonumber \\[2mm]
J^{PC}=2^{++}: \quad 
& X_{2} = \lvert 1_{\de}, 1_{\bde} \rangle_{2} \, ,
\end{align}
where the numbers inside (outside) each ket denote $s_{\de,\bde}$ ($S$), respectively.

This notation originates with the earlier diquark-antidiquark picture of Refs.~\cite{Maiani:2004vq,Maiani:2014aja}, as does the assertion that the largest spin couplings in the state are those between quarks within $\de$ or $\bde$ [as indicated by the $\kappa$ term in Eq.~(\ref{eq:FullHam}) below].  The dynamical diquark model differs in ($i$) explaining the separation of the $\de\bde$ pair as due to the generation of a finite-size color flux tube connecting them, which in its static limit can be described using the BO formalism; and ($ii$) by including isospin-dependent (and other) operators.

$S$-wave tetraquarks also serve as the natural starting point for constructing orbital excitations~\cite{Lebed:2017min}.  These states arise when the $\de\bde$ relative orbital angular momentum $L$ is nonzero.  In this work we focus on the first such excitation, the $P$-wave 
($L=1$) sector, whose states form the BO multiplet $\Sigma^+_g(1P)$.  Such states necessarily have negative parity, $P=(-1)^L=-1$, and 
display a richer structure than their $S$-wave counterparts because the total angular momentum $\mathbf{J} \equiv \mathbf{L}+\mathbf{S}$ can assume a variety of values.   Moreover, the accompanying charge-conjugation quantum number $C$ depends upon both $L$ and $S$, allowing the appearance of both conventional and genuinely exotic (for a $q\bar q$ meson) $J^{PC}$ assignments.  In particular, the $P$-wave tetraquark sector naturally accommodates quantum numbers such as $1^{-+}$ and $0^{--}$, which cannot arise from a simple $q\bar q$ system.

To classify these states systematically, we construct the $P$-wave basis kets
\[
\lvert s_{\de}, s_{\bde} \rangle^{L=1}_{S},
\]
from which all possible total $J^{PC}$ quantum numbers in the multiplet $\Sigma^+_g(1P)$ follow.  The complete set of $P$-wave configurations in this multiplet, including their diquark spin couplings and $J^{PC}$ assignments (and adapted from Ref.~\cite{Lebed:2017min}), is listed in Table~\ref{tab:tetra_spins}\@.   This classification provides the foundation for our analysis of $P = -$ hidden-strangeness tetraquarks in the following sections.

\begin{table}[t]
\centering
\caption{$P$-wave $\de\bde$ states in the basis $|s_{\de},s_\bde\rangle_{S}^{L=1}$ for the multiplet $\Sigma^+_g(1P)$.  The subscripts $\de$,$\, \bde$ inside the kets and the superscripts $L=1$ are suppressed for clarity.  The letter symbol in each case ($Z^\prime$, $X_2$, {\it etc.}) indicates the core $\de \bde$ state [Eq.~(\ref{eq:Swave})] that combines with $L=1$ to produce the given state, with total $J$ indicated by the superscript.}
\renewcommand{\arraystretch}{1.4}
\begin{tabular*}{0.85\columnwidth}{c@{\extracolsep{\fill}}l@{\extracolsep{\fill}}}
\hline\hline
$J^{PC}$   & State  \\
\hline
$0^{-+}$  & $Z_{P}^{(0)}=\frac{1}{\sqrt{2}}\big(|1,0\rangle_{ 1} - |0,1\rangle_{ 1}\big)$ \\
$0^{-+}$ &   $Z_{P}^{\prime(0)}=|1,1\rangle_1$ \\
 
$0^{--}$ &   $X_{1P}^{(0)}=\frac{1}{\sqrt{2}}\big(|1,0\rangle_{ 1} + |0,1\rangle_{ 1}\big)$\   \\

$1^{-+}$ & $Z_{P}^{(1)}=\frac{1}{\sqrt{2}}\big(|1,0\rangle_{ 1} - |1,0\rangle_{ 1}\big)$ \\
$1^{-+}$ & $Z_{P}^{\prime(1)}=|1,1\rangle_1$ \\

$1^{--}$ &   $X_{0P}^{ (1)}=|0,0\rangle_{ 0}$ \\
$1^{--}$ &   $X_{0P}^{\prime(1)}=|1,1\rangle_0$ \\
$1^{--}$   & $X_{1P}^{(1)}=\frac{1}{\sqrt{2}}\big(|1,0\rangle_{ 1} + |0,1\rangle_{ 1}\big)$ \\
$1^{--}$ &   $X_{2P}^{(1)}=|1,1\rangle_2$   \\

$2^{-+}$   & $Z_{P}^{(2)}=\frac{1}{\sqrt{2}}\big(|1,0\rangle_{ 1} - |0,1\rangle_{ 1}\big)$ \\
$2^{-+}$ &   $Z_{P}^{\prime(2)}=|1,1\rangle_1$   \\
 
$2^{--}$    & $X_{1P}^{(2)}=\frac{1}{\sqrt{2}}(|1,0\rangle_{ 1} + |0,1\rangle_{ 1}) $   \\
$2^{--}$    & $X_{2P}^{(2)}=|1,1\rangle_2$ \\

$3^{--}$ &   $X_{2P}^{(3)}=|1,1\rangle_2$   \\
\hline\hline
\end{tabular*}
\label{tab:tetra_spins}
\end{table}

\section{Mass Spectrum of $P$-Wave Hidden-Strangeness Tetraquarks}
\label{sec:MassSpec}

To describe $P$-wave tetraquarks in the dynamical-diquark framework, we use the following Hamiltonian \cite{Giron:2020fvd}:
\begin{align}
H &= H_0+ 2\kappa_{qs}\!\left({\bf s}_q\!\cdot\!{\bf s}_s
+ {\bf s}_{\bar q}\!\cdot\!{\bf s}_{\bar s}\right)
+ V_{LS}\,{\bf L}\!\cdot\!{\bf S}
 \nonumber \\
&\quad
+ V_I\,{\bm\tau}_q\!\cdot\!{\bm\tau}_{\bar q}\,
{\bm\sigma}_q\!\cdot\!{\bm\sigma}_{\bar q}
+ V_T\,{\bm\tau}_q\!\cdot\!{\bm\tau}_{\bar q}\,
S_{12}^{(q\bar q)} .
\label{eq:FullHam}
\end{align}
The term $H_0$ encodes the spin-averaged mass of the $P$-wave tetraquark
multiplet $\Sigma^+_g(1P)$; all remaining contributions generate fine-structure mass splittings.  
The coefficient $\kappa_{qs}$ controls the intra-diquark hyperfine
interactions, while $V_{LS}$ governs the coupling between orbital and total
quark spin.  
The isospin–spin coupling proportional to $V_I$ and the tensor interaction
weighted by $V_T$ further refine the spectrum.  The tensor operator used above is defined for any two constituent spins
${\bf S}_1$ and ${\bf S}_2$ with separation vector {\bf r} as:
\begin{align}
S_{12} \equiv 4\left[
3\,\frac{({\bf S}_1\!\cdot\!{\bf r})({\bf S}_2\!\cdot\!{\bf r})}{r^2}
- {\bf S}_1\!\cdot\!{\bf S}_2
\right],
\label{eq:Tensor}
\end{align}
which allows its application either to just the light $q\bar q$ pair or to
the $\de\bde$ degrees of freedom.  For the explicit forms of the mass terms for each quantum number, see Eqs.~(21)-(34) of Ref.~\cite{Giron:2020fvd}.

The eigenvalue of $H_0$ for the $P$-wave $s\bar s q\bar q$ tetraquark multiplet $\Sigma^+_g(1P)$ in the
dynamical diquark model~\cite{Jafarzade:2025qvx} is predicted to be $M_0(1P)=2.311~\text{GeV}$, and represents the
mass of the multiplet prior to the inclusion of fine-structure interactions.
Once the spin-spin, spin-orbit, isospin, and tensor terms are incorporated,
the degeneracy of the multiplet is lifted, and the individual states are split
by approximately $100$-$150~\mathrm{MeV}$, consistent with the magnitude estimated
in Ref.~\cite{Giron:2020fvd}.  The interactions of Eq.~(\ref{eq:FullHam}) therefore delimit the
characteristic mass window in which the members of this multiplet are expected
to appear.

Guided by the set of allowed $J^{PC}$ quantum numbers (Table~\ref{tab:tetra_spins}), we focus upon 6 experimentally observed resonances listed
by the PDG~\cite{ParticleDataGroup:2024cfk} that fall within this interval:
\[
\rho(2150), \, \phi(2170), \, \omega(2220), \,
\eta(2225), \, \rho_3(2250), \, \eta(2370) \, .
\]
For the isoscalar sector, we use PDG average masses for $\phi(2170)$, $\eta(2225)$, and $\eta(2370)$.  For the isovector states $\rho(2150)$ and $\rho_3(2250)$, however, no reliable PDG averages exist; therefore, we adopt BESIII measurements obtained through the decay channel $\psi(2S)\to K^+K^-\eta$ \cite{BESIII:2022yzp}.  Indeed, beyond the fact that this channel exhibits manifest
strange–quark production, we also find that it  generates a fit quality that is
significantly better than the other decay channels do when they are tested under
the tetraquark hypothesis.

The model parameters are determined through a least-squares fit by minimizing a $\chi^2$ function weighted by experimental uncertainties.  The resulting reduced chi-square is $\chi^2_{\rm red} \equiv \chi^2/\mathrm{d.o.f.}=0.94$, and the best-fit parameter values are
\begin{align}
\kappa_{qs} & = \ \ 19.75  \pm 4.65 ~\mathrm{MeV}\,, \nonumber \\
V_{LS}      & = -38.79   \pm 4.08 ~\mathrm{MeV}\,, \nonumber \\
V_I         & = \ \ \ \; 3.63 \pm 1.17~\mathrm{MeV} \,, \nonumber \\
V_T         & = -16.27  \pm 0.77~\mathrm{MeV} \,.
\label{eq:FitParams}
\end{align}

The experimental masses, fitted values, and corresponding 
$\chi^2_{\text{red}}$ contributions are summarized in 
Table~\ref{tab:fit}\@.  
Among all channels, $\omega(2220)$ contributes most significantly to $\chi^2_{\rm red}$.  The PDG quotes a world-average mass of 
$2188 \pm 21~\text{MeV}$, based upon three BESIII measurements:
$2153 \pm 30$~\cite{BESIII:2024qjv}, 
$2176 \pm 24$~\cite{BESIII:2020xmw}, and 
$2232 \pm 19$~\cite{BESIII:2022yzp}.  
If the lightest $I=0$, $1^{--}$ tetraquark is identified with
$\phi(2170)$, then the only BESIII value that provides a reasonable identification with the next-heaviest $I=0$, $1^{--}$ state within the tetraquark multiplet is 
the higher-mass determination from Ref.~\cite{BESIII:2022yzp}.   If the lower values contributing to the PDG fit are used, then $\chi^2_{\rm red}$ rises from 0.94 to 4.01, rendering the identification of $\omega(2220)$ as a tetraquark state untenable in this model.  Thus, one may take the larger fit value as a model prediction that the higher-mass $\omega(2220)$ state exists, and that the lower mass measurements will either be deprecated through future BESIII measurements, or indicate the presence of an additional lower-mass state with non-tetraquark structure. 
 \begin{table}[h]
   \centering
\caption{Fit to the $P$-wave hidden-strangeness tetraquark resonances within the dynamical diquark model.  The table lists the assigned states, their experimental masses, the corresponding theoretical model predictions, and the resulting $\chi^2$ contributions.}
\renewcommand{\arraystretch}{1.2}
\begin{tabular*}{\columnwidth}{c@{\extracolsep{\fill}}cccc}
\hline\hline
States & $I^G\,(J^{PC})$ & Mass (Exp.) & Mass (Th.)  & $\chi^2$ \\
\hline
$\rho(2150)$ & $1^+\,(1^{--})$ & $2255^{+17}_{-18}$ \cite{BESIII:2019dme} & $2248\pm 3$ & $0.15$   \\
 
$\phi(2170)$ & $0^-\,(1^{--})$ & $2164\pm 5$ \cite{ParticleDataGroup:2024cfk} & $2163\pm 3$ & $0.03$   \\

$\omega(2220)$ & $0^-\,(1^{--})$ & $2232\pm 19$ \cite{BESIII:2022yzp} & $2257\pm13$ & $1.18$  \\
 
$\eta(2225)$ & $0^+\,(0^{-+})$ & $2221^{+13}_{-11}$ \cite{ParticleDataGroup:2024cfk} & $2223\pm12$ & $0.01$   \\
 
$\rho_3(2250)$ & $1^+\,(3^{--})$ & $2248\pm 17$ \cite{BESIII:2019dme} & $2263\pm14$ & $0.46$   \\
 
$\eta(2370)$ & $0^+\,(0^{-+})$ & $2377\pm 9$ \cite{ParticleDataGroup:2024cfk} & $2380\pm13$ & $0.04$   \\
\hline\hline
\end{tabular*}
\label{tab:fit}
\end{table}

The negative-parity states below 2~GeV, such as $\rho(1900)$, $\rho_3(1990)$, and  $\pi_2(2005)$ lie at least $200~\mathrm{MeV}$ below $M_0(1P)$.  Accommodating these states would require mass shifts well beyond those generated by spin-dependent effects alone.  They are therefore excluded from the fit and are likely influenced by additional dynamics, such as mixing with excited conventional $q\bar q$ states or coupled-channel effects.  Note that we do not consider the mixing between the tetraquark states, for the sake of simplicity in this analysis.

\begin{figure*}
    \centering
    \includegraphics[scale=0.7]{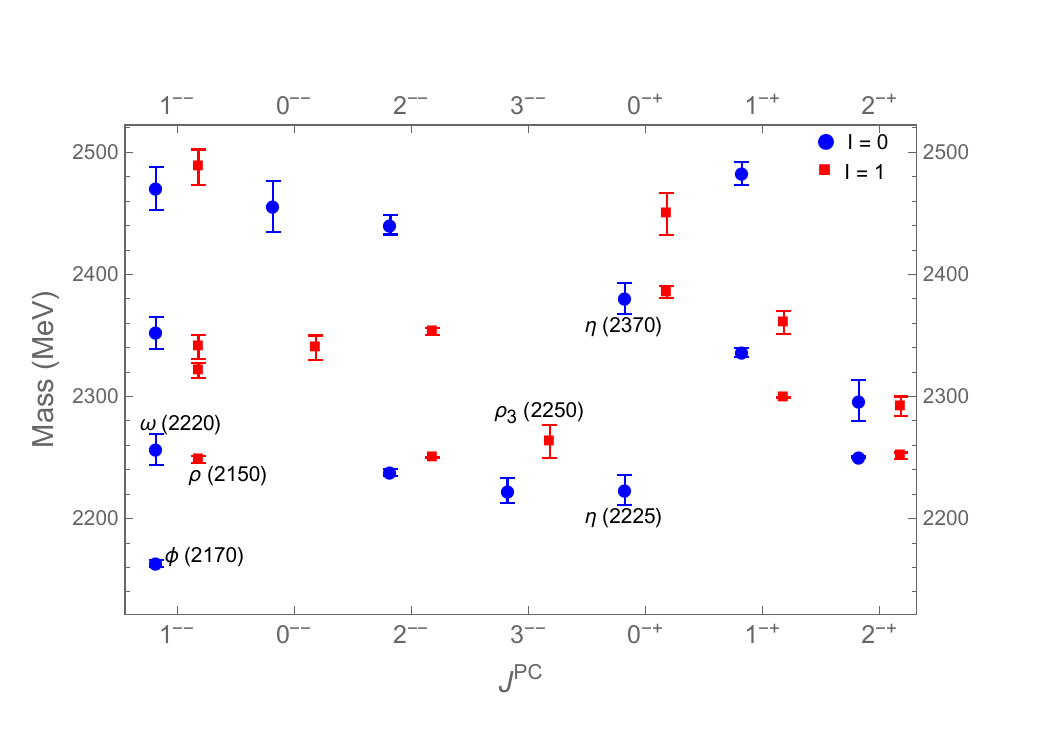}
    \caption{Predicted spectrum of the $P$-wave hidden-strangeness tetraquarks comprising the $\Sigma^+_g(1P)$ multiplet of the dynamical diquark model.  Blue (circle) and red (square) points denote $I=0$ ($I=1)$ states, respectively.  States used as fit inputs are labeled.}
    \label{fig:spectrum}
\end{figure*}

Model parameters fixed by the fit [Eq.~(\ref{eq:FitParams})] are used to generate the full $P$-wave hidden-strangeness tetraquark multiplet $\Sigma^+_g(1P)$ spectrum in Table \ref{tab:AllMasses}, which is organized according
to state quantum numbers $J^{PC}$ and
isospin $I$, with separate columns for the $I=0$ and $I=1$ sectors (see also Fig.~\ref{fig:spectrum}).
Note the state multiplicities: a single pair ($I=0$ and $I=1$) for $0^{--}$ and $3^{--}$, 4 pairs for $1^{--}$, and 2 pairs for all other $J^{PC}$.
The quoted uncertainties arise from the propagation of the model parameters
in Eq.~(\ref{eq:FitParams}), and provide an estimate of the theoretical spread of the predictions.

A notable feature of the spectrum is the proliferation of states
in the mass region between $2.2$ and $2.4~\mathrm{GeV}$, which substantially overlaps
with the energy range in which a number of experimentally observed
resonances have been reported but not yet confirmed (see Table \ref{tab:further-states}). Nevertheless, we predict some exotic isoscalar states like $\phi_0(2455)$ and $\eta_1(2480)$  to lie above 2.4 GeV\@.

In the $1^{--}$ sector, the lowest predicted isoscalar masses near
$2.16$ and $2.26~\mathrm{GeV}$ naturally accommodate the experimentally
observed $\phi(2170)$ and $\omega(2220)$, while the corresponding
isovector state around $2.25~\mathrm{GeV}$ aligns with the mass for
$\rho(2150)$ observed by BESIII \cite{BESIII:2019dme}.  All other states with $J^{PC} = 1^{--}$ are predicted to lie above
$2.3~\mathrm{GeV}$, suggesting the existence of additional vector
resonances that have not yet been clearly established experimentally.  Resonances that are compatible with our predictions are the isovector $\rho(2270)$~\cite{Anisovich:2002su,OmegaPhoton:1985vyt} and the isoscalars $\omega(2290)$~\cite{Bugg:2004rj} and $\omega(2330)$~\cite{OmegaPhoton:1988guj}, listed in Table~\ref{tab:further-states} [although they cease to be suitable hidden-strangeness tetraquark candidates if $\rho(2150)$ and $\omega(2220)$ fill those roles].

In the  $0^{-+}$ sector, we predict an isoscalar state
near $2.22~\mathrm{GeV}$ compatible with $\eta(2225)$, along with
a heavier partner around $2.38~\mathrm{GeV}$ that may be associated
with $\eta(2370)$.  On the other hand, Table~\ref{tab:further-states} contains observed but unconfirmed resonances of masses broadly compatible with our predictions,  such as $\eta(2190)$~\cite{Bugg:1999jc} and $\eta(2320)$~\cite{Anisovich:2000ix}.

Similar mass
hierarchies between the isoscalar states appear in the $1^{-+}$ and  $2^{--}$ channels.  While current data offers no candidate with the mass and the quantum numbers matching to our predictions for $1^{-+}$, there are $2^{--}$ isoscalar $\omega_2(2195)$ and isovector $\rho_2(2225)$ states that match our predictions well but need further confirmation.  Note that well-established ground-state mesons with quantum numbers
$2^{--}$ are also absent in the conventional spectrum~\cite{Jafarzade:2022uqo},
which further complicates the identification of tetraquark states with these
quantum numbers.

The exotic $0^{--}$ quantum number cannot be accommodated within the
conventional $q\bar q$ meson framework, and therefore provides a
particularly clean probe of multiquark dynamics.  In our tetraquark
spectrum, an isoscalar and an isovector $0^{--}$ state  are predicted in the mass region
$2.3$-$2.5~\mathrm{GeV}$, with a representative isoscalar mass around
$2.45~\mathrm{GeV}$.  At present, no firmly established experimental
candidate with these quantum numbers exists, and the PDG does not list
any confirmed $0^{--}$ light-meson states.  Nevertheless, the predicted
mass lies in a region accessible to current high-statistics experiments
such as COMPASS and BESIII, where searches for multi-meson and
strange-rich final states could provide critical tests of the tetraquark
interpretation.

A candidate $2^{-+}$ state $\pi_{2}(2100)$, with a reported decay
width of $620\pm50~\mathrm{MeV}$, has been observed primarily in the
$f_{2}\pi$ channel.  An early measurement by the ACCMOR
Collaboration~\cite{ACCMOR:1980llh} reported a mass of
$2100\pm150~\mathrm{MeV}$, which is compatible (within uncertainties)
with our predicted mass of about $2.25~\mathrm{GeV}$\@.  The large width
may be attributed in part to the $S$-wave nature of the $f_{2}\pi$
decay, as we discuss in the next section.  Nevertheless, an analysis of three decades ago~\cite{VES:1995mdc} obtains a mass of $2090\pm 30$~MeV, rather lower than our prediction for $s\bar s q\bar q$ states.  The PDG currently
lists this state with the remark ``needs confirmation''; hence we have not included this state in the fit procedure.  Other possible candidates (according to mass and $J^{PC}$) are the isoscalar $\eta_2(2250)$ and isovector $\pi_2(2285)$, which also need confirmation.  Recent phenomenological analyses in Ref.~\cite{Farina:2023oqk} predict $2^{-+}$ hybrid states to lie in the $2.3$-$2.4~\mathrm{GeV}$ region.

Finally, the presence of a $J^{PC}=3^{--}$ isovector state at
$2.26~\mathrm{GeV}$, used as input for the $\rho_{3}(2250)$, demonstrates
that higher-spin resonances can be naturally accommodated within the same
tetraquark framework.  Other unconfirmed spin-3 states observed in the predicted mass region include $\omega_3(2255)$~\cite{Anisovich:2002xoo} and $\omega_3(2285)$~\cite{Bugg:2004rj,Anisovich:2002xoo}.

 Compared to the hybrid model of Refs.~\cite{Swanson:2023zlm,Farina:2023oqk}, the predicted mass eigenvalue for
the isoscalar $1^{--}$ state is similar in both pictures (and, of course, $q\bar q g$ hybrids would have suppressed decays into strange mesons).  For the remaining quantum numbers, however, the two frameworks predict clearly separated mass regions: the hybrid model places isoscalar $1^{-+}$ and $2^{-+}$ states at $1750$--$1850$ and $1650$--$1800$~MeV, respectively, well below the tetraquark predictions of $2340$--$2480$ and
$2250$--$2300$~MeV, while for $0^{-+}$ the ordering reverses, with the hybrid model
predicting $2430$--$2560$~MeV, versus the tetraquark prediction of
$2220$--$2380$~MeV. 

Overall, the predicted spectrum exhibits a coherent
structure across isospin and $J^{PC}$ sectors, simultaneously
accounting for several experimentally observed states and predicting
additional partners that may have been seen but are yet unconfirmed, and are likely accessible in future experimental
studies.

\begin{table*}
\caption{The complete predicted $\Sigma^+_g(1P)$ tetraquark spectrum.  Boldface entries denote the postdicted states that are employed as inputs to the fit: $\eta(2225),\ \eta(2370),\ \phi(2170),\ \omega(2220),\ \rho(2150),\ \rho_3(2250)$, respectively.}
\label{tab:AllMasses}
\centering

\setlength{\extrarowheight}{0.5ex}
\begin{tabular}{c c c @{ \ } c c}
\hline\hline
$J^{PC}$ & \multicolumn{2}{c}{$I=0$} & \multicolumn{2}{c}{$I=1$} \\
\hline
$0^{-+}$ & $\bf{2223.5\pm 12.4}$ & $\bf{2380.2\pm 12.6}$ & $2385.4\pm 4.8$ & $2449.6\pm 16.9$  \\

$0^{--}$ & $2455.6\pm 20.9$ & & $2339.9\pm 10.1$\\

$1^{-+}$ & $2336.20\pm3.7$ & $2482.8\pm 9.5$ & $2299.2\pm 0.4$ & $2360.6\pm 9.4$  \\

$1^{--}$ & $\bf{2163.4\pm 2.8}$& $\bf{2256.5\pm 12.7}$   & $\bf{2248.5\pm 2.6}$ & $2321.5\pm 6.0$ \\
& $2352.3\pm 13.1$ & $2470.3\pm 17.8$ & $2340.8\pm 9.7$ &  $2488.0\pm 14.4$  \\ 
             
$2^{-+}$ & $2250.2\pm 0.8$ & $2296.5\pm 16.8$  & $2251.7\pm 2.5$ & $2292\pm 8.0$ \\

$2^{--}$ & $2237.7\pm 2.8$ & $2440.6\pm 7.9$ & $2250.0\pm 0.5$ &  $2353.2\pm 3.1$  \\

 $3^{--}$ & $2222.8\pm 10.2$ & & $\bf{2263.3\pm 13.7}$\\  
\hline\hline

\end{tabular}

\end{table*}

\begin{table*}[t]
 \centering
\caption{Negative-parity light mesons with mass above 2.1~GeV listed as ``Further States'' in the PDG\@.  
Masses and total widths $\Gamma$ (in MeV) are taken from the experimental analyses cited.}
\label{tab:further-states}
 
\begin{tabular}{c@{\hskip 1em}c@{\hskip 1em}c@{\hskip 1em}c@{\hskip 1em}c}
\hline\hline
State & $I^G(J^{PC})$ & {Mass (MeV)} & {$\Gamma$ (MeV)} & Reference(s) \\ \hline

$\eta(2190)$     
& $0^+(0^{-+})$   
& $2190\pm 50$            
& $850\pm 100$            
&  \cite{Bugg:1999jc} \\

$\omega_2(2195)$ 
& $0^-(2^{--})$   
& $2195\pm 30$            
& $225\pm 40$             
&  \cite{Anisovich:2002xoo} \\

$\rho_2(2225)$   
& $1^+(2^{--})$   
& $2225\pm 35$            
& $335^{+100}_{-50}$             
& \cite{Anisovich:2002su} \\

$\eta_2(2250)$   
& $0^+(2^{-+})$   
& $2248\pm20$ / $2267\pm 14$ 
& $280\pm 20$ / $290\pm 50$ 
& \cite{Anisovich:2000us} \!/\! \cite{Anisovich:2000ut} \\

$\omega_3(2255)$ 
& $0^-(3^{--})$   
& $2255\pm 15$            
& $175\pm 30$             
&  \cite{Anisovich:2002xoo} \\

$\rho(2270)$     
& $1^+(1^{--})$   
& $2265\pm 40$ / $2280\pm 50 $
& $325\pm 80$ / $440\pm 110$
& \cite{Anisovich:2002su} \!/\! \cite{OmegaPhoton:1985vyt}\\

$\pi_2(2285)$    
& $1^-(2^{-+})$   
& $2285^{+20}_{-25}$      
& $250^{+20}_{-25}$       
& \cite{Anisovich:2010nh} \\

$\omega_3(2285)$ 
& $0^-(3^{--})$   
& $2278\pm 28$ / $2285\pm 60$ 
& $224\pm 50$ / $230\pm 40$ 
& \cite{Bugg:2004rj} \!/\! \cite{Anisovich:2002xoo} \\

$\omega(2290)$   
& $0^-(1^{--})$   
& $2290\pm 20$            
& $275\pm 35$             
&\cite{Bugg:2004rj} \\

$\eta(2320)$     
& $0^+(0^{-+})$   
& $2320\pm 15$            
& $230\pm 35$             
& \cite{Anisovich:2000ix} \\

$\omega(2330)$   
& $0^-(1^{--})$   
& $2330\pm 30$            
& $435\pm 75$             
& \cite{OmegaPhoton:1988guj} \\
\hline\hline
\end{tabular}
\end{table*}

\section{Fall-Apart Decays of $P$-Wave Tetraquarks}

In this section we present the results for the fall-apart (valence-quark rearrangement) decays of the $P$-wave tetraquarks.  For the fully strange tetraquarks, fall-apart decays have been discussed ({\it e.g.}, Refs.~\cite{Ke:2018evd,Liu:2020lpw,Liu:2026ljb}).  Nevertheless, we derive the fall-apart decay widths of the $P$-wave tetraquarks in this section.

In the fall-apart  mechanism, the initial tetraquark state
\(\mathcal{T}\) with total angular momentum \(J\) and mass \(M_{\mathcal T}\)
rearranges into two color-singlet mesons \(M_1\) and \(M_2\) with spins
\(J_1,J_2\) and masses \(m_1,m_2\).  The two-body partial width is given in the
parent rest frame by the standard expression
\begin{equation}\label{eq:width_general}
\Gamma_{ \mathcal{T}\to M_1 M_2}
=\frac{\mathcal{P}}{4\pi (2J+1)}
\sum_{\text{spins}} \bigl| \mathcal{M}_{fi}\bigr|^2,
\end{equation}
where \(\mathcal{M}_{fi}\) is the transition amplitude, the phase-space factor \(\mathcal{P}\equiv \frac{k}{8\pi M_{\mathcal T}^2}\), and  \(k\equiv|\mathbf{k}|\)  is the magnitude of the three-momentum of either
final-state meson:
\begin{equation}
k=\frac{1}{2M_{\mathcal T}}\sqrt{\bigl[M_{\mathcal T}^2-(m_1+m_2)^2\bigr]
\bigl[M_{\mathcal T}^2-(m_1-m_2)^2\bigr]} .
\label{eq:CMmomentum}
\end{equation}

The mesons listed in Table~\ref{tab:meson_qn} represent the primary hadronic
degrees of freedom that appear as decay products of hidden-strangeness
tetraquarks.  In the fall-apart decay mechanism, the tetraquark rearranges
into two color-singlet $q\bar q$ (including $s$ quarks) mesons without the need for additional
quark-antiquark pair creation.  Consequently, the relevant final states
consist of the light pseudoscalar ($P$), vector ($V$), axial-vector
($A$ [$1^{++}$] and $B$ [$1^{+-}$]), and tensor ($T$) mesons.  Table~\ref{tab:meson_qn} summarizes the
quark-model quantum numbers of these mesons: the orbital angular momentum
of the $q\bar q'$ pair $L_{q\bar q'}$, the total quark spin $S_{q\bar q'}$,
and the resulting meson spin $J_M$.  The scalar states $f_0$, $a_0$ are not included, as they have been argued to already contain a $\de\bde$ component~\cite{Jaffe:1976ig}.  The quoted masses correspond to the
central values reported by the PDG\@.  These states
form the building blocks for the di-meson channels into which
hidden-strangeness tetraquarks may decay.

\begin{table*}
\centering
\caption{Meson states, their quark-model quantum numbers, and  measured masses $M_{\rm PDG}$, rounded to the nearest MeV\@.
$L_{q\bar q'}$ is the $q\bar q'$ orbital angular momentum, 
$S_{q\bar q'}$ is the $q\bar q'$ spin, and $J_M$ is the meson total spin.}
\renewcommand{\arraystretch}{1.2}
\begin{tabular}{c c c c c}
\hline\hline
State & $L_{q\bar q'}$ & $S_{q\bar q'}$ & $J_M$ & $M_{\rm PDG}$ [MeV] \\
\hline
$P\equiv\{\pi,\,K,\,\eta,\,\eta'\}$ 
& 0 & 0 & 0 
& $\{140,\;494,\;548,\;958\}$ \\

$V\equiv\{\rho,\,K^{\ast},\,\omega,\,\phi\}$ 
& 0 & 1 & 1 
& $\{775,\;890,\;783,\;1020\}$ \\

$B\equiv\{b_{1},\,K_{1B},\,h_{1},\,h_{1}'\}$ 
& 1 & 0 & 1 
& $\{1230,\;1253/1403,\;1166, 1409\}$ \\

$A\equiv\{a_{1},\,K_{1A},\,f_{1},\,f_{1}'\}$ 
& 1 & 1 & 1 
& $\{1230,\;1253/1403,\;1282,\;1428\}$ \\

$T\equiv\{a_{2},\,K_{2}^{\ast},\,f_{2},\,f_{2}'\}$ 
& 1 & 1 & 2 
& $\{1318,\;1427,\;1275,\; 1517\}$ \\
\hline\hline
\end{tabular}
\label{tab:meson_qn}
\end{table*}

Among the pseudoscalar mesons, the physical $\eta$ and $\eta'$ are not
pure flavor states, but rather arise from the mixing of the light-quark and
strange-quark configurations.  In the quark-flavor basis, one defines
\begin{align}
|\eta_q\rangle &= 
\frac{1}{\sqrt{2}}\left(|u\bar u\rangle+|d\bar d\rangle\right), \\
|\eta_s\rangle &= |s\bar s\rangle ,
\end{align}
and the physical states are obtained through a single mixing angle
$\phi_{P}$,
\begin{align}
|\eta\rangle &= 
\cos\phi_{P}\,|\eta_q\rangle 
- \sin\phi_{P}\,|\eta_s\rangle , \nonumber \\
|\eta'\rangle &= 
\sin\phi_{P}\,|\eta_q\rangle 
+ \cos\phi_{P}\,|\eta_s\rangle .
\label{eq:0--mix}
\end{align}
This mixing is
particularly important for hidden-strangeness tetraquark decays, because
the $s\bar s$ component of the pseudoscalar states can couple directly
to the strange quarks present in the tetraquark configuration.

For the remaining isoscalar mesons shown in
Table~\ref{tab:meson_qn}, such as $\omega$-$\phi$ in the $V$ sector
and $f_2 \,$-$f_2'$ in the $T$ sector,
flavor mixing is known to occur but is close to ideal, while in the axial ($A$) and pseudovector ($B$) sector the magnitude of the mixing is not settled \cite{Giacosa:2023fdz,Giacosa:2024epf}.
In the present analysis, we therefore neglect these mixings and treat
the states as approximately pure $q\bar q$  ($h_1, f_1, f_2$) and $s\bar s$  ($h'_1, f'_1, f'_2$) configurations.
The only exception we explicitly include is the pseudoscalar
$\eta$-$\eta'$ system, whose mixing plays a significant role in decay
amplitudes involving hidden-strangeness components.  This approximation
simplifies the construction of decay channels, while retaining the most
phenomenologically important mixing effects relevant for
hidden-strangeness tetraquark decays.

 We also consider  mixing between the axial ($A$) and pseudovector ($B$) kaons.  In the
quark model these states have $J^P=1^+$: the
${}^{3}P_{1}$ state ($K_{1A}$) and the ${}^{1}P_{1}$ state ($K_{1B}$).
The physical mesons $K_1(1270)$ and $K_1(1400)$ arise from the mixing
of these two basis states:
\begin{align}\nonumber
|K_1(1270)\rangle &=
\sin\theta_K\,|K_{1A}\rangle
+\cos\theta_K\,|K_{1B}\rangle,\\
|K_1(1400)\rangle &=
\cos\theta_K\,|K_{1A}\rangle
-\sin\theta_K\,|K_{1B}\rangle ,
\label{eq:Kmix}
\end{align}
where $\theta_K$ is the kaonic mixing angle.  Because
of this mixing, both $K_1(1270)$ and $K_1(1400)$ appear in the decay
channels associated with the $A$ and $B$ multiplets listed in Table \ref{tab:meson_qn}.

\subsection{Spin-Recoupling Factor}

The angular-momentum structure of the fall-apart decay amplitude can be expressed
in a model-independent form, using standard recoupling
coefficients.  The relevant transformation between different coupling schemes
of 4 angular momenta is governed by the Wigner $9j$ symbol.  In general one
has
\begin{equation}
\begin{aligned}
\bigl|(j_1 j_2) j_{12},& (j_3 j_4) j_{34}; J\bigr\rangle
=
\sqrt{[j_{12}][j_{34}][j_{13}][j_{24}]}
\\
&\quad\times
\begin{Bmatrix}
j_1 & j_2 & j_{12} \\
j_3 & j_4 & j_{34} \\
j_{13} & j_{24} & J
\end{Bmatrix}
\bigl|(j_1 j_3) j_{13}, (j_2 j_4) j_{24}; J\bigr\rangle ,
\end{aligned}
\end{equation}
using the shorthand notation $[j]\equiv 2j+1$.

For hidden-strangeness tetraquarks, we identify the 4 constituent spins as
\begin{equation}
{\bf j}_1 \to {\bf s}_q, \qquad
{\bf j}_2 \to {\bf s}_{\bar q}, \qquad
{\bf j}_3 \to {\bf s}_s, \qquad
{\bf j}_4 \to {\bf s}_{\bar s}. 
\end{equation}
The intermediate couplings relevant for the meson and diquark bases are
\begin{equation}
{\bf j}_{12} \to {\bf s}_{q\bar q}, \ \;
{\bf j}_{34} \to {\bf s}_{s\bar s}, \ \;
{\bf j}_{13} \to {\bf s}_{(q s) \, = \, \de}, \ \;
{\bf j}_{24} \to {\bf s}_{(\bar q \bar s) \, = \, \bde} \, , 
\end{equation}
and $J \to S$ denotes the total angular momentum carried by all 4 valence quarks.

The overlap between the $\de\bde$ spin configuration
$(qs)(\bar q\bar s)$ and the di-meson spin basis
$(q\bar q)(s\bar s)$ is therefore
\begin{equation}
\begin{aligned}
&\Bigl\langle
(s_q s_{\bar q}) s_{q\bar q},\,
(s_s s_{\bar s}) s_{s\bar s}; S \,
\Big|
(s_q s_s) s_{\delta},\,
(s_{\bar q} s_{\bar s}) s_{\bar\delta}; S
\Bigr\rangle
\\
&\qquad =
\sqrt{[S_{q\bar q}][S_{s\bar s}][s_{\delta}][s_{\bar\delta}]}
\begin{Bmatrix}
s_q & s_{\bar q} & s_{q\bar q} \\
s_s & s_{\bar s} & s_{s\bar s} \\
s_{\delta} & s_{\bar\delta} & S
\end{Bmatrix}.
\end{aligned}
\end{equation}
If the final-state mesons instead carry kaonic di-meson quantum numbers $(q \bar s)(\bar q s)$ rather than $(q\bar q)(s\bar s)$, then the same expressions throughout this section continue to hold, upon the global substitution of all subscripts $\bar q \leftrightarrow \bar s$.

The expression above describes the spin recoupling relevant for
$S$-wave tetraquark decays.  For higher excitations, the orbital
degrees of freedom must also be included.  The tetraquark and the final
di-meson system carry a relative orbital angular momentum $L$:
\begin{equation} 
{\bf J} = {\bf S} + {\bf L} . 
\end{equation}

The total angular momenta of the two final-state mesons are obtained by coupling
their constituent spins and orbital angular momenta,
\begin{align}
{\bf J}_{M_1} = {\bf s}_{q\bar q} + {\bf L}_{q\bar q}, \qquad
{\bf J}_{M_2} = {\bf s}_{s\bar s} + {\bf L}_{s\bar s},
\end{align}
while the total angular momentum of the  di-meson system (in its $S$ partial wave) satisfies
\begin{equation}
{\bf J} = {\bf J}_{M_1} + {\bf J}_{M_2}.
\end{equation}

The recoupling between the $(SL)$ and $(J)$ bases is again expressed
through a Wigner $9j$ symbol,
\begin{equation}
\begin{aligned}
&\Bigl\langle
(s_{q\bar q} L_{q\bar q}) J_{M_1},
(s_{s\bar s} L_{s\bar s}) J_{M_2};\, J \,
\Big|
(s_{q\bar q} s_{s\bar s}) S,
(L_{q\bar q} L_{s\bar s}) L;\, J
\Bigr\rangle
\\
&\qquad =
\sqrt{[J_{M_1}][J_{M_2}][S][L]}
\begin{Bmatrix}
s_{q\bar q} & L_{q\bar q} & J_{M_1} \\
s_{s\bar s} & L_{s\bar s} & J_{M_2} \\
S & L & J
\end{Bmatrix}.
\end{aligned}
\end{equation}

Combining the spin and orbital recoupling factors, the complete
spin-orbital recoupling factor entering the fall-apart decay amplitude can be
written as
\begin{equation}
\begin{aligned}
\mathcal S & \equiv
\sqrt{[s_{q\bar q}][s_{s\bar s}][s_{\delta}][s_{\bar\delta}]}
\begin{Bmatrix}
s_q & s_{\bar q} & s_{q\bar q} \\
s_s & s_{\bar s} & s_{s\bar s} \\
s_{\delta} & s_{\bar\delta} & S
\end{Bmatrix}
\\
&\quad\times
\sqrt{[J_{M_1}][J_{M_2}][S][L]}
\begin{Bmatrix}
s_{q\bar q} & L_{q\bar q} & J_{M_1} \\
s_{s\bar s} & L_{s\bar s} & J_{M_2} \\
S & L & J
\end{Bmatrix}.
\end{aligned}
\label{eq:SpinFactor}
\end{equation}

For the case of interest in this work, namely $P$-wave tetraquarks,
the relative orbital angular momentum is $L=1$.  The factor $\mathcal S$
therefore represents the complete spin-orbital recoupling factor
that multiplies the spatial overlap in the fall-apart decay amplitude.

\subsection{Decay Channels}

Table~\ref{tab:orbital_L} lists all allowed values of the relative
orbital angular momentum $\ell$ for fall-apart decays of tetraquarks with
given $J^{PC}$ quantum numbers into pairs of mesons $M_1 M_2$ (or $MM$ if they are identical).  The selection
rules follow from angular-momentum conservation, together with parity and
charge-conjugation constraints, and determine which decay channels can
proceed in low partial waves and therefore dominate the observed widths.

For final states consisting of two identical pseudoscalar mesons ($PP$), the allowed
decays are highly constrained.  Only tetraquarks with $J^{PC}=1^{--}$ can
decay through a $P$-wave ($\ell=1$), while higher-spin states such as
$J^{PC}=3^{--}$ require an $F$-wave ($\ell=3$).  Consequently, fall-apart
decays into $PP$ channels are generally suppressed. For nonidentical pseudoscalar pairs $P_1 P_2$ such
as $\eta\eta'$, a $P$-wave decay is also allowed for the exotic channel $1^{-+}$.

Decays into pseudoscalar-vector ($PV$) final states are phenomenologically
important for vector tetraquarks.  A $P$-wave fall-apart decay is allowed for
$1^{--}$ states, and is consistent with observed channels such as
$\phi(2170)\to\phi\eta^{(\prime)}$, $\omega(2220)\to\omega\eta^{(\prime)}$, and $\rho(2150)\to\omega\pi^0$.

Final states containing two vector mesons ($VV$ or $V_1V_2$) are generally
allowed for all $\Sigma^+_g(1P)$ tetraquark quantum numbers, with the exception of the exotic $0^{--}$ state into two identical vectors such as $\phi\phi$.  Several partial waves may contribute, most commonly $\ell=1$ and $3$, while $\ell = 5$ becomes possible for $J^{PC} = 3^{--}$.  The experimentally observed decay  $\eta(2225)\to\phi\phi$~\cite{BESIII:2016qzq} provides an ($\ell = 1$) example of such a channel, whereas the kaonic decay $K^\ast\bar K^\ast$ ($V_1 V_2$) has not yet been observed with statistical significance for $\phi(2170)$~\cite{BES:2009rue}.

Channels involving axial-vector ($A$), pseudovector ($B$), or tensor ($T$) mesons exhibit a richer structure.  Mixed final states, such as
$PA$, $PB$, and $PT$, allow even values of $\ell$, including $\ell=0$ for
certain $J^{PC}$ assignments.  These $S$-wave decays are particularly
important phenomenologically, because they lead to unsuppressed amplitudes
near threshold, and are expected to dominate the total decay width when kinematically
allowed.  For example, the exotic state $1^{-+}$ can decay into $PA$ and $PB$
through an $S$-wave fall-apart process, whereas the exotic $0^{--}$ state is
forbidden from decaying into these channels in any partial wave.  Both $1^{-+}$ and $0^{--}$ states can nevertheless decay into $PV$ through a $P$-wave.  The observed $2^{-+}$ decay $\pi_2(2100)\to f_2\pi$ provides an $S$-wave $PT$ example.

The remaining light di-meson channels relevant for fall-apart decays are
$VA$, $VB$, and $VT$.  In the $VA$ and $VB$ systems, all tetraquark states
except $3^{--}$ can decay through an $S$-wave, while the $3^{--}$ state
requires higher partial waves.  In contrast, the $VT$ channel permits an
$S$-wave decay  for all channels except $0^{-+}$ and $0^{--}$.

Finally, baryon-antibaryon channels such as $\Lambda\bar{\Lambda}$ and
$\Xi\bar{\Xi}$ (strictly not fall-apart modes, and for $\Xi\bar{\Xi}$ are more relevant for $s\bar s s\bar s$ states) are allowed for all $J^{PC}$ assignments except the exotic quantum numbers $0^{--}$ and $1^{-+}$.  Among $\Lambda\bar{\Lambda}$ and
$\Xi\bar{\Xi}$, $S$-wave decays are allowed only for $0^{-+}$ and $1^{--}$ tetraquarks.

\begin{table}[h!]
\centering
\renewcommand{\arraystretch}{1.2}
\caption{Allowed orbital angular momenta $\ell$ (denoting partial waves) for each decay channel and for given tetraquark $J^{PC}$, where repeated ($MM$) and distinct ($M_1 M_2$) symbols stand for identical and nonidentical meson pairs, respectively.}
\begin{tabular*}{\columnwidth}{c@{\extracolsep{\fill}}ccccccc}
\hline\hline
& \multicolumn{7}{c}{$\ell$ for $J^{PC}$} \\
\cline{2-8} 
Channel 
& $0^{-+}$ & $0^{--}$ & $1^{-+}$ & $1^{--}$ & $2^{-+}$ &  $2^{--}$ &     $3^{--}$ \\
\hline
$PP$          & - &  - & - & 1  &-  &-  &3  \\
$P_1P_2$          & - &  - & 1 & 1  &-  &-  &3  \\
$VV$          & 1 & - & 1 &  1,3& 1,3  &1,3  &1,3,5  \\
$V_1V_2$          & 1 & 1 & 1,3 &  1,3& 1,3  &1,3  &1,3,5  \\
$AA$          & 1 &-  &1  &1,3  &1,3  &1,3  & 1,3,5 \\
$A_1A_2$          & 1 &1  &1,3  &1,3  &1,3  &1,3  & 1,3,5 \\
$BB$          & 1  & -  & 1 & 1,3 & 1,3  &1,3  & 1,3,5 \\
$B_1B_2$          & 1  & 1  & 1,3 & 1,3 & 1,3  &1,3  & 1,3,5 \\
$PV$          &1  & 1 & 1 & 1 &1,3   &1,3  & 3 \\
$PA$          & - & - & 0,2 & 0,2  & 2 & 2 & 2,4 \\
$PB$          &-  & - &0,2  & 0,2 & 2 & 2  & 2,4  \\
$PT$          & 2  & 2  & 2 & 2  & 0,2,4  & 0,2,4  & 2,4 \\
$VA$          & 0,2  & 0,2 & 0,2 & 0,2  & 0,2,4 & 0,2,4 & 2,4  \\
$VB$          &  0,2  & 0,2 & 0,2 & 0,2  & 0,2,4 & 0,2,4 & 2,4 \\
$AB$ & 1  & 1  & 1,3 & 1,3 & 1,3  &1,3  & 1,3,5 \\
 $VT$ & 2  & 2  & 0,2,4 & 0,2,4  & 0,2,4  & 0,2,4  & 0,2,4,6 \\
 $AT$ & 1,3 & 1,3 & 1,3 & 1,3 & 1,3,5  &1,3,5  & 1,3,5\\
 $BT$ & 1,3 & 1,3 & 1,3 & 1,3 & 1,3,5  &1,3,5  & 1,3,5\\
 $TT$ & 1,3 & -  & 1,3 & 1,3,5 & 1,3,5  &1,3,5  & 1,3,5,7\\
 $T_1T_2$ & 1,3 & 1,3  & 1,3,5 & 1,3,5 & 1,3,5  &1,3,5  & 1,3,5,7\\
 {$\Lambda\bar{\Lambda}$} & 0 & -  & -  & 0,2 & 2& 2&2,4\\
 {$\Xi\bar{\Xi}$} & 0 &-  &  - & 0,2 & 2& 2&2,4\\
\hline\hline
\end{tabular*}
\label{tab:orbital_L}
\end{table}

\subsection{Numerical Results for the Decay Ratios}

We estimate relative decay-width ratios for $S$-wave fall-apart channels of
hidden–strangeness $P$-wave tetraquarks using the phase-space factors $\mathcal P$
and spin-recoupling factors $\mathcal S$ computed in Tables~VII–XII.
Throughout this analysis, we take the mixing angles
$\phi_P \simeq 39^\circ$ [Eqs.~(\ref{eq:0--mix})] and $\theta_K \simeq 33^\circ$ [Eqs.~(\ref{eq:Kmix})]~\cite{ParticleDataGroup:2024cfk,Cheng:2011pb}, and assume that the partial widths scale as
\begin{equation}
\Gamma_i \propto \mathcal P_i |\mathcal S_i|^2 .
\end{equation}
This approximation, equivalent to assuming that all 2-body $S$-wave fall-apart decays for a given tetraquark have the same invariant amplitude ${\cal M}_{ fi}$ apart from spin-recoupling factors ${\cal S}_{ i}$, enables us to identify the dominant decay channels for each
tetraquark multiplet and to compare the predicted decay patterns for the
experimentally observed  and predicted resonances.

\subsubsection{$0^{-+}$ Tetraquarks}

The decay ratios obtained from Table~\ref{tab:0mp_dominant} for the
state $Z_P^{(0)}$ can be compared for the isoscalar resonances
$\eta(2225)$ and $\eta(2370)$.  Here and below, we present results in each $J^{PC}$ sector for the  specific tetraquark state that has the most nonzero spin-recoupling factors ${\cal S}$.  In the present case, this choice means using $Z^{(0)}_P$ rather than $Z^{\prime \, (0)}_P$.

For the heavier candidate $\eta(2370)$, the predicted  decay-width ratios, where a combination such as $K_1(1270)\bar K^\ast$ is understood to include its charge-conjugate, are
\begin{align*}
& \Gamma_{K_1(1270)\bar K^\ast}
:\Gamma_{f_1'\omega}
:\Gamma_{f_1\phi}
:\Gamma_{h_1\phi}
:\Gamma_{h'_1\omega}
\\
\simeq \, &
1.00:0.64:0.45:0.35:0.35 \, .
\end{align*}
For the lighter $\eta(2225)$ state, several channels become more
suppressed (or even kinematically forbidden) relative to the dominant mode:
\begin{align*}
& \Gamma_{K_1(1270)\bar K^\ast}
:\Gamma_{f_1'\omega}
:\Gamma_{h_1\phi}
:\Gamma_{h_1'\omega}
\\ \simeq \, &
1.00:0.30:0.30:0.23 \, .
\end{align*}
For the lighter isovector partner $0^{-+}$ $\pi(2385)$, we obtain
\begin{align*}
& \Gamma_{K_1(1270)\bar K^\ast}
:\Gamma_{f_1' \rho}
:\Gamma_{a_1\phi}
:\Gamma_{h_1' \rho}
:\Gamma_{b_1\phi}
\\ \simeq \, &
1.00:0.65:0.59:0.34:0.29 \, .
\end{align*}
Using the same spin–recoupling factors for the heavier isovector state $\pi(2450)$, the decay-width ratios can be estimated by rescaling with the corresponding phase-space factors, which gives
\begin{align*}
& \Gamma_{K_1(1270)\bar K^\ast}
:\Gamma_{f_1' \rho}
:\Gamma_{a_1\phi}
:\Gamma_{h_1' \rho}
:\Gamma_{b_1\phi}
\\ \simeq \, &
1.00:0.66:0.62:0.35:0.31 \, .
\end{align*}
The overall decay pattern remains the same across both isovector
states, again highlighting the importance of the
$K_1(1270)\bar K^\ast$ mode. The competing mode $K_1(1400)\bar K^\ast$ turns out to be strongly suppressed, reflecting destructive interference in the  $\theta_K$-dependent factor of the spin-recoupling factor ${\cal S}$.

\subsubsection{$0^{--}$ Tetraquarks}

Using the phase–space and spin–coupling factors listed in
Table~\ref{tab:0mm_dominant}, we find the ratios for dominant decays of the
isoscalar state $\phi_0(2455)$ to be
\begin{align*}
& \Gamma_{K_1(1270)\bar K^\ast}
:\Gamma_{f_1'\omega}
:\Gamma_{f_1\phi}
:\Gamma_{h_1\phi}
:\Gamma_{h_1'\omega}
\\ \simeq \, &
1.00:0.68:0.56:0.35:0.35 \, .
\end{align*}
The corresponding ratios for the $0^{--}$ isovector partner $\rho_0(2340)$ are
\begin{align*}
& \Gamma_{K_1(1270)\bar K^\ast}
:\Gamma_{f_1' \rho}
:\Gamma_{a_1\phi}
:\Gamma_{h_1' \rho}
:\Gamma_{b_1\phi}
\\
\simeq \, &
1.00:0.63:0.53:0.34:0.26 \, .
\end{align*}
In both cases, the decay pattern is clearly dominated by
$K_1(1270)\bar K^\ast$ , with $K_1(1400) \bar K^\ast$ again heavily suppressed due to destructive interference in its ${\cal S}$ factor.

\subsubsection{$1^{-+}$ Tetraquarks}

For the exotic state $\eta_1(2340)$ [chosen to correspond to $Z_P^{(1)}]$ and using the results of Table~\ref{tab:1mp_dominant}, we find the
predicted decay-width ratios to be
\begin{align*}
& \Gamma_{K_1(1270) \bar K^\ast}
:\Gamma_{K_1(1400) \bar K}
:\Gamma_{h_1\phi}
:\Gamma_{h_1'\omega}
:\Gamma_{f_1' \eta} \! :\\
& \Gamma_{f_1 \eta}
:\Gamma_{K_1(1270) \bar K}
:\Gamma_{f_1 \eta'}
:\Gamma_{f_1'\omega}
:\Gamma_{f_1\phi}
\\ \simeq \, &
1.00:0.65:0.61:0.57:0.51: \\
& 0.40:0.32:0.29:0.26:0.14 \, .
\end{align*}
For $\eta_1(2480)$, we obtain
\begin{align*}
& \Gamma_{K_1(1270) \bar K^\ast}
:\Gamma_{h_1 \phi}
:\Gamma_{h_1'\omega}
:\Gamma_{K_1(1400) \bar K}
:\Gamma_{f_1' \eta}
:\Gamma_{f_1 \eta} \! : \\
& \Gamma_{f_1 \eta'}
:\Gamma_{f_1'\omega}
:\Gamma_{K_1(1270) \bar K}
:\Gamma_{f_1 \phi}
:\Gamma_{f_1' \eta'}
\\ \simeq \, &
1.00:0.63:0.60:0.56:0.45:0.30:
\\ & 0.34:0.29:0.27:0.25:0.15 \, .
\end{align*}
For the isovector partners $\pi_1(2300)$ and $\pi_1(2360)$, the dominant
ratios are, respectively,
\begin{align*}
& \Gamma_{f_1' \pi}
:\Gamma_{K_1(1270) \bar K^\ast}
:\Gamma_{K_1(1400) \bar K}
:\Gamma_{h_1' \rho}
:\Gamma_{a_1 \eta}: \\
& \Gamma_{b_1 \phi}
:\Gamma_{K_1(1270) \bar K}
:\Gamma_{a_1 \eta'}
:\Gamma_{f_1' \rho}
:\Gamma_{a_1 \phi}
\\ \simeq \, &
1.00:0.89:0.61:0.50:0.40: \\
& 0.35:0.30:0.30:0.23:0.17 \, ,
\end{align*}
and
\begin{align*}
& \Gamma_{f_1' \pi}
:\Gamma_{K_1(1270) \bar K^\ast}
:\Gamma_{\bar K_1(1400) \bar K}
:\Gamma_{h_1' \rho}
:\Gamma_{b_1 \phi} \! : \\
& \Gamma_{a_1 \eta}
:\Gamma_{a_1 \eta'}
:\Gamma_{K_1(1270) \bar K}
:\Gamma_{f_1' \rho}
:\Gamma_{a_1 \phi}
\\ \simeq \,
& 1.00:0.99:0.61:0.58:0.47: \\
& 0.40:0.35:0.30:0.27:0.24 \, .
\end{align*}
In the $1^{-+}$ case, the decay patterns support
several channels contributing with comparable strength.  The dominant
modes are $K_1(1270) \bar K^\ast$ for the isoscalar states
$\eta_1(2340)$ and $\eta_1(2480)$, while $f_1(1420) \pi$  is slightly favored for
decays of the isovector states $\pi_1(2300)$ and $\pi_1(2360)$.

\subsubsection{$1^{--}$ Tetraquarks}

Using the phase-space and spin-recoupling factors in Table~\ref{tab:1mm_dominant},
we find the ratios for dominant decays of $\phi(2170)$ [chosen to correspond to $X_{1P}^{(1)}]$ to be
\begin{align*}
& \Gamma_{K_1(1400) \bar K}
:\Gamma_{f_1' \eta}
:\Gamma_{K_1(1270) \bar K^\ast}
:\Gamma_{f_1 \eta}
:\Gamma_{K_1(1270) \bar K}
\\ \simeq \, &
1.00:0.73:0.65:0.64:0.54 \, .
\end{align*}
The corresponding results for $\omega(2220)$ are
\begin{align*}
& \Gamma_{K_1(1270) \bar K^\ast}
:\Gamma_{K_1(1400) \bar K}
:\Gamma_{f_1' \eta}
:\Gamma_{h_1 \phi}:\\
& \Gamma_{h_1'\omega}
:\Gamma_{f_1 \eta}
:\Gamma_{K_1(1270) \bar K}
:\Gamma_{f_1'\omega}
:\Gamma_{f_1 \eta'}
\\\simeq \, &
1.00:0.77:0.58:0.54:0.49:0.48:0.39:0.21:0.16 \, .
\end{align*}
For the lightest $1^{--}$ isovector partner, which we identify with $\rho(2150)$, the dominant ratios are
\begin{align*}
& \Gamma_{f_1' \pi}
:\Gamma_{K_1(1270) \bar K^\ast}
:\Gamma_{K_1(1400) \bar K}
:\Gamma_{a_1 \eta}
:\Gamma_{h_1' \rho}: \\
& \Gamma_{K_1(1270) \bar K}
:\Gamma_{a_1 \eta'}
:\Gamma_{f_1' \rho}
\\ \simeq \, &
1.00:0.76:0.59:0.39:0.39:0.30:0.23:0.16 \, .
\end{align*}
Here, for the first time, we find the dominant fall-apart mode to be non-kaonic.

\subsubsection{$2^{-+}$ Tetraquarks}

For the lightest  isoscalar $2^{-+}$ tetraquark $\eta_2(2250)$  [identified
with $Z_P^{(2)}$], using the results of Table~\ref{tab:2mp_dominant} we obtain
\begin{align*}
& \Gamma_{K_2^\ast \bar K} :
\Gamma_{K_1(1270) \bar K^\ast} :
\Gamma_{h_1\phi} :
\Gamma_{f_2 \eta} \! : \\
& \Gamma_{f_2' \eta} :
\Gamma_{h_1'\omega} :
\Gamma_{f_1'\omega} :
\Gamma_{f_2 \eta'}
\\ \simeq \, &
1.00 : 0.95 : 0.50 : 0.46 : 0.46 : 0.46 : 0.19 : 0.15 .
\end{align*}
For the isovector state $\pi_2(2250)$, the ratios become
\begin{align*}
& \Gamma_{f_2' \pi} :
\Gamma_{K_2^\ast \bar K} :
\Gamma_{K_1(1270) \bar K^\ast} :
\Gamma_{h_1' \rho} :
\Gamma_{a_2 \eta} :
\Gamma_{f_1' \rho}
\\ \simeq \, &
1.00 : 0.89 : 0.85 : 0.44 : 0.39 : 0.19 .
\end{align*}
In the isovector sector, the dominant contribution arises from the
channel $f_2' \pi$.  The observation of the $f_2\pi$ decay mode for
$\pi_2(2100)$ suggests that tensor–pseudoscalar channels may
play an important role in the decay dynamics of the $2^{-+}$ isovector
states.

The heavier $\eta_2(2300)$ and $\pi_2(2290)$ states are so similar in mass to their lighter partners that one expects nearly identical respective decay patterns.

\subsubsection{ $2^{--}$ Tetraquarks}

For the $2^{--}$ isoscalar tetraquark $\phi_2(2240)$ [chosen to correspond to $X_{1P}^{(2)}$] and using the results of Table~\ref{tab:2mm_dominant}, we find the predicted decay-width ratios to be
\begin{align*}
& \Gamma_{K_2^\ast \bar K}
:\Gamma_{K_1(1270) \bar K^\ast}
:\Gamma_{f_2 \eta}
:\Gamma_{h_1 \phi} \! : \\
& \Gamma_{f_2' \eta}
:\Gamma_{h_1'\omega}
:\Gamma_{f_1'\omega}
:\Gamma_{f_2 \eta'}
\\ \simeq \, &
1.00:0.91:0.47:0.46:0.46:0.41:0.16:0.09 \, .
\end{align*}
For its lighter isovector partner $\rho_2(2250)$, we obtain
\begin{align*}
& \Gamma_{f_2' \pi}
:\Gamma_{K_2^\ast \bar K}
:\Gamma_{K_1(1270)\bar K^\ast}
:\Gamma_{h_1' \rho}
:\Gamma_{a_2 \eta}
:\Gamma_{f_1' \rho}
\\ \simeq \, &
1.00:0.89:0.86:0.43:0.41:0.18 \, .
\end{align*}
For the heavier states $\phi_2(2440)$ and $\rho_2(2350)$, the dominant
channels remain similar:
\begin{align*}
& \Gamma_{K_1(1270)\bar K^\ast}
:\Gamma_{K_2^\ast \bar K}
:\Gamma_{f_2\phi}
:\Gamma_{f'_2 \omega}
: \Gamma_{K_2^\ast \bar K^\ast}
:\Gamma_{h_1 \phi} \! : \\
& \Gamma_{h'_1 \omega}
:\Gamma_{f'_2 \eta}
:\Gamma_{f_2 \eta}
:\Gamma_{f_2 \eta'}
:\Gamma_{f'_1 \omega}
:\Gamma_{f_1 \phi}
\\ \simeq \,
& 1.00:0.79:0.72:0.67:0.64:0.62: \\
& 0.60:0.41:0.35:0.34:0.28:0.23 \, ,
\end{align*}
and
\begin{align*}
& \Gamma_{K_1(1270)\bar K^\ast}
:\Gamma_{f'_2 \pi}
:\Gamma_{K_2^\ast \bar K}
:\Gamma_{h'_1 \rho}
:\Gamma_{f'_2 \rho}
:\Gamma_{b_1 \phi} \! : \\
& \Gamma_{K_2^\ast(1430) \bar K^\ast}
:\Gamma_{a_2 \eta}
:\Gamma_{f'_1 \rho}
:\Gamma_{a_2 \phi}
:\Gamma_{a_2 \eta'}
:\Gamma_{a_1 \phi}
\\ \simeq \,
& 1.00:0.96:0.88:0.59:0.53:0.49: \\
& 0.41:0.36:0.28:0.27:0.25:0.25 \, ,
\end{align*}
respectively.  The leading, but not truly dominant, channels are therefore $K_2^\ast(1430) \bar K$ for $\phi_2(2240)$,
$K_1(1270)\bar K^\ast$ for $\phi_2(2440)$,
$f_2(1525) \pi$ for $\rho_2(2250)$,
and $K_1(1270)\bar K^\ast$ for $\rho_2(2350)$.

\subsubsection{$3^{--}$ Tetraquarks}

 The $3^{--}$ states $\omega_3(2220)$ (predicted) and $\rho_3(2250)$ (observed) from Table~\ref{tab:AllMasses}, if considered as $s\bar s q\bar q$ tetraquarks, are too light to undergo fall-apart $S$-wave decays into 2-meson states taken from Table~\ref{tab:meson_qn}. Hence, unlike all the other $\Sigma^+_g(1P)$ states, there are no relevant decay channels to tabulate.

This is not to say that the $3^{--}$ states are especially narrow.  According to Table~\ref{tab:orbital_L}, they allow $P$-wave decays for $VV$ and multiple other 2-meson combinations.  Moreover, if non-fall-apart OZI-suppressed decays of $s\bar s q\bar q$ are permitted, then channels such as $N\bar N$ become possible; this particular mode has actually been seen for $\rho_3(2250)$, although not in recent years~\cite{ParticleDataGroup:2024cfk}.

As discussed in Sec.~\ref{sec:MassSpec}, either of two unconfirmed $\omega_3$ states  [$\omega_3(2255)$ and $\omega_3(2285)$ in Table~\ref{tab:further-states}] could serve as the predicted partner to $\rho_3(2250)$ in the $\Sigma^+_g(1P)$ multiplet of the dynamical diquark model.

\begin{table*}[ht!]
\caption{Dominant fall-apart decay channels of the lowest $J^{PC}=0^{-+}$ tetraquark states ($I=0,1$).  All listed modes proceed in a relative $S$ wave
($\ell=0$).  ${\cal S}$ is the spin factor given in Eq.~(\ref{eq:SpinFactor}), and $\bar{\cal P}$ is $10^3$ times the phase-space factor ${\cal P}$ (in GeV$^{-1}$) given above Eq.~(\ref{eq:CMmomentum}).}
\label{tab:0mp_dominant}
\centering
\renewcommand{\arraystretch}{1.2}
\begin{tabular}{c l @{\hskip 1em} c @{\hskip 1em} c c c}
\hline\hline
Isospin & Decay channel & $\mathcal{S}$($Z_P^{(0)}$) & $\mathcal{S}$($Z_P^{\prime(0)}$) & $\bar{\mathcal{P}}$[$\eta(2225)$] & $\bar{\mathcal{P}}$[$\eta(2370)$] \\
\hline
$I=0$
& $f_1(1420)\,\omega(782)$                        & $\frac{1}{\sqrt{2}}$ & - & 0.9 & 3.0 \\
& $f_1(1285)\,\phi(1020)$                          &  $\frac{1}{\sqrt{2}}$ & -& 0 & 2.1 \\
& $h_1(1170)\,\phi(1020)$                          &  $\frac{1}{2}$ & $\frac{1}{\sqrt{2}}$&  1.6 & 3.3 \\
& $h_1(1415)\,\omega(782)$                        & $\frac{1}{2}$ & $\frac{1}{\sqrt{2}}$ & 1.4 & 3.1\\
& $K_1(1270)\,\bar K^{*}(892)+\mathrm{c.c.}$ & $\frac{\sqrt{2}\sin{\theta_K}+\cos{\theta_K}}{2}$ & $\frac{\cos{\theta_k}}{\sqrt{2}}$   & 2.3 & 3.6 \\
& $K_1(1400)\,\bar K^{*}(892)+\mathrm{c.c.}$ & $\frac{\cos{\theta_K}-\sqrt{2}\sin{\theta_K}}{2}$ &  $-\frac{\sin{\theta_K}}{\sqrt{2}}$ & 0 & 2.2\\
\hline
 Isospin & Decay channel & $\mathcal{S}$ ($Z_P^{(0)}$) & $\mathcal{S}$ ($Z_P^{\prime(0)}$) & $\bar{\mathcal{P}}$[$\pi(2385)$] & $\bar{\mathcal{P}}$[$\pi(2450)$] \\\hline
$I=1$
& $a_1(1260)\,\phi(1020) $                      & $\frac{1}{\sqrt{2}}$ & - & 2.8 & 3.2\\
& $b_1(1235)\,\phi(1020) $                      & $\frac{1}{2}$ & $\frac{1}{\sqrt{2}}$ & 2.8 & 3.2  \\
& $\rho(770)\,h_1(1415) $                      & $\frac{1}{2}$ & $\frac{1}{\sqrt{2}}$ & 3.2 & 3.6 \\
& $\rho(770)\,f_1(1420) $                      & $\frac{1}{\sqrt{2}}$ & - & 3.1 & 3.4 \\
& $K_1(1270)\,\bar K^{*}(892)+\mathrm{c.c.}$ & $\frac{\sqrt{2}\sin{\theta_K}+\cos{\theta_K}}{2}$ & $\frac{\cos{\theta_K}}{\sqrt{2}}$  & 3.6 & 3.9\\
& $K_1(1400)\,\bar K^{*}(892)+\mathrm{c.c.}$ & $\frac{\cos{\theta_K}-\sqrt{2}\sin{\theta_K}}{2}$ & $-\frac{\sin{\theta_K}}{\sqrt{2}}$  & 2.2  &2.8 \\\hline\hline
\end{tabular}
 
\end{table*}

\begin{table*}[ht!]
\caption{Dominant fall-apart decay channels of the lowest $J^{PC}=0^{--}$ tetraquark states ($I=0,1$).  All listed modes proceed in a relative $S$ wave
($\ell=0$).  ${\cal S}$ is the spin factor given in Eq.~(\ref{eq:SpinFactor}), and $\bar{\cal P}$ is $10^3$ times the phase-space factor ${\cal P}$ (in GeV$^{-1}$) given above Eq.~(\ref{eq:CMmomentum}).}
\label{tab:0mm_dominant}
\centering
\renewcommand{\arraystretch}{1.2}
\begin{tabular}{c l @{\hskip 1em} c c}
\hline\hline
Isospin & Decay channel & $\mathcal{S}(X_{1P}^{(0)})$   & $\bar{\mathcal{P}}[\phi_0(2455)]$ \\
\hline
$I=0$
& $f_1(1420)\,\omega(782)$                        & $\frac{1}{\sqrt{2}}$ & 3.4  \\
& $f_1(1285)\,\phi(1020)$                          &  $\frac{1}{\sqrt{2}}$& 2.8 \\
& $h_1(1170)\,\phi(1020)$                          & $\frac{1}{2}$ &3.7  \\
& $h_1(1415)\,\omega(782)$                        & $\frac{1}{2}$ & 3.5 \\
& $K_1(1270)\,\bar K^{*}(892)+\mathrm{c.c.}$ & $\frac{\sqrt{2}\sin{\theta_K}+\cos{\theta_K}}{2}$ & 3.9 \\
& $K_1(1400)\,\bar K^{*}(892)+\mathrm{c.c.}$ &  $\frac{-\sqrt{2}\sin{\theta_K}+\cos{\theta_K}}{2}$&2.8  \\
\hline
Isospin & Decay channel & $\mathcal{S}(X_{1P}^{(0)})$   & $\bar{\mathcal{P}}[\rho_0(2340)]$ \\
\hline
$I=1$
& $a_1(1260)\,\phi(1020) $                      & $\frac{1}{\sqrt{2}}$ & 2.3 \\
& $b_1(1235)\,\phi(1020) $                      &$\frac{1}{2}$  &2.3  \\
& $\rho(770)\,h_1(1415) $                      & $\frac{1}{2}$ &2.9  \\
& $\rho(770)\,f_1(1420) $                      & $\frac{1}{\sqrt{2}}$ & 2.8 \\
& $K_1(1270)\,\bar K^{*}(892)+\mathrm{c.c.}$ & $\frac{\sqrt{2}\sin{\theta_K}+\cos{\theta_K}}{2}$ & 3.4 \\
& $K_1(1400)\,\bar K^{*}(892)+\mathrm{c.c.}$ &$\frac{-\sqrt{2}\sin{\theta_K}+\cos{\theta_K}}{2}$  & 1.7 \\\hline\hline
\end{tabular} 
\end{table*}

\begin{table*}
\caption{Dominant fall-apart decay channels of the lowest $J^{PC}=1^{-+}$ tetraquark states ($I=0,1$).  All listed modes proceed in a relative $S$ wave
($\ell=0$).  ${\cal S}$ is the spin factor given in Eq.~(\ref{eq:SpinFactor}), and $\bar{\cal P}$ is $10^3$ times the phase-space factor ${\cal P}$ (in GeV$^{-1}$) given above Eq.~(\ref{eq:CMmomentum}).}
\label{tab:1mp_dominant}
\centering
\renewcommand{\arraystretch}{1.2}
\begin{tabular}{c c @{\hskip 1em} c @{\hskip 1em} l @{\hskip 1em} c c }
\hline\hline
Isospin & Decay channel & $\mathcal{S}$ ($Z_P^{(1)}$) & $\mathcal{S}$ ($Z_P^{\prime(1)}$) & $\bar{\mathcal{P}}$[$\eta_1(2340)$] & $\bar{\mathcal{P}}$[$\eta_1(2480)$] \\
\hline
$I=0$
& $\eta(548)\,f_1(1285)$                          & $\frac{-\sin{\phi_P}}{2}$ & $\frac{\sin{\phi_P}}{\sqrt{2}}$ &  5.0 & 5.2 \\
& $\eta(548)\,f_1(1420)$                          & $\frac{\cos{\phi_P}}{2}$ & $\frac{\cos{\phi_P}}{\sqrt{2}}$ & 4.2 & 4.5 \\
& $\eta(548)\,h_1(1170)$                          & - &- & 5.6 & 5.6 \\
& $\eta(548)\,h_1(1415)$                          & - & - & 4.3 & 4.6 \\
& $\eta(958)\,f_1(1285)$                          & $\frac{\cos{\phi_P}}{2}$ & $\frac{\cos{\phi_P}}{\sqrt{2}}$ & 2.4 & 3.4 \\
& $\eta(958)\,f_1(1420)$                          & $\frac{\sin{\phi_P}}{2}$ & $\frac{\sin{\phi_P}}{\sqrt{2}}$ & 0.0 & 2.2 \\
& $\eta(958)\,h_1(1170)$                          & - & - & 3.5 & 4.1 \\
& $\eta(958)\,h_1(1415)$                          & - & - & 0.0 & 2.4 \\
& $f_1(1420)\,\omega(782)$                        & $\frac{1}{2\sqrt{2}}$ & - & 2.6 & 3.5 \\
& $f_1(1285)\,\phi(1020)$                          & $\frac{1}{2\sqrt{2}}$ &- & 1.4 & 3.0 \\
& $h_1(1170)\,\phi(1020)$                          & $\frac{1}{2}$ & $\frac{1}{\sqrt{2}}$ & 3.0 & 3.8 \\
& $h_1(1415)\,\omega(782)$                        &  $\frac{1}{2}$ & $\frac{1}{\sqrt{2}}$ & 2.8 & 3.6\\
& $K\, \bar K_1(1270)+\mathrm{c.c.}$               & $\frac{\sin{\theta_K}}{2}$ &  $\frac{\sin{\theta_K} }{\sqrt{2}}$   & 5.3 & 5.4 \\
& $K\, \bar K_1(1400)+\mathrm{c.c.}$               & $\frac{\cos{\theta_K}}{2}$ & $\frac{\cos{\theta_K}}{\sqrt{2}}$  & 4.6 & 4.8 \\
& $K^\star(782)\, \bar K_1(1270)+\mathrm{c.c.}$               & $\frac{\sin{\theta_K}}{2\sqrt{2}}+\frac{\cos{\theta_K}}{2}$ & $\frac{\cos{\theta_K}}{\sqrt{2}}$  &3.3 & 4.0  \\
& $K^\star(782)\, \bar K_1(1400)+\mathrm{c.c.}$               & $\frac{\cos{\theta_K}}{2\sqrt{2}}-\frac{\sin{\theta_K}}{2} $ & $\frac{-\sin{\theta_K}}{\sqrt{2}}$  & 1.6 & 3.0\\\hline
  Isospin & Decay channel & $\mathcal{S}$ ($Z_P^{(1)}$) & $\mathcal{S}$ ($Z_P^{\prime(1)}$) & $\bar{\mathcal{P}}$[$\pi_1(2300)$] & $\bar{\mathcal{P}}$[$\pi_1(2360)$] \\\hline

$I=1$
& $\eta(548)\,a_1(1260)$                           & $\frac{-\sin{\phi_P}}{2}$ & $\frac{-\sin{\phi_P}}{\sqrt{2}}$ & 5.2 & 5.3 \\
& $\eta(958)\,a_1(1260)$                           &  $\frac{\cos{\phi_P}}{2}$ & $\frac{\cos{\phi_P}}{\sqrt{2}}$ &2.6 & 3.1 \\
& $\eta(548)\,b_1(1235)$                           & -  & - & 5.2& 5.3 \\
& $\eta(958)\,b_1(1235)$                           &  -& - &2.6& 3.1  \\
& $\pi\,f_1(1420)$                          & $\frac{1}{2}$ & $\frac{1}{\sqrt{2}}$ & 5.2 & 5.3 \\
& $\pi\,h_1(1415)$                          &  -  & - & 5.3 & 5.4 \\
& $\phi(1020)\,a_1(1260)$                          &  $\frac{1}{2\sqrt{2}}$ & - & 1.8 & 2.5 \\
& $\phi(1020)\,b_1(1235)$                          & $\frac{1}{2}$ & $\frac{1}{\sqrt{2}}$ &  1.8 & 2.5\\
& $\rho(770)\,h_1(1415) $                      & $\frac{1}{2}$ & $\frac{1}{\sqrt{2}}$ &2.6 & 3.1  \\
& $\rho(770)\,f_1(1420) $                      &  $\frac{1}{2\sqrt{2}}$ & - &2.4 & 2.9  \\
& $K\, \bar K_1(1270)+\mathrm{c.c.}$              & $\frac{\sin{\theta_K} }{2}$ &  $\frac{\sin{\theta_K}}{\sqrt{2}}$   & 5.3 & 5.4 \\
& $K\, \bar K_1(1400)+\mathrm{c.c.}$               & $\frac{ \cos{\theta_K}}{2}$ & $\frac{\cos{\theta_K}}{\sqrt{2}}$  &4.5 & 4.6\\
& $K^\star(782)\, \bar K_1(1270)+\mathrm{c.c.}$               & $\frac{\sin{\theta_K}}{2\sqrt{2}}+\frac{\cos{\theta_K}}{2}$ & $\frac{\cos{\theta_K}}{\sqrt{2}}$  &   3.1 & 3.5  \\
& $K^\star(782)\, \bar K_1(1400)+\mathrm{c.c.}$                & $\frac{\cos{\theta_K}}{2\sqrt{2}}-\frac{\sin{\theta_K}}{2 }$ & $\frac{-\sin{\theta_K}}{\sqrt{2}}$  & 0.6 & 2.0  \\
\hline\hline
\end{tabular}
\end{table*}

\begin{table*}
\caption{Dominant fall-apart decay channels of the lowest $J^{PC}=1^{--}$ tetraquark states ($I=0,1$).  All listed modes proceed in  a relative $S$ wave
 ($\ell=0$).  ${\cal S}$ is the spin factor given in Eq.~(\ref{eq:SpinFactor}), and $\bar{\cal P}$ is $10^3$ times the phase-space factor ${\cal P}$ (in GeV$^{-1}$) given above Eq.~(\ref{eq:CMmomentum}).}
\label{tab:1mm_dominant}
\centering
\renewcommand{\arraystretch}{1.2}
\begin{tabular}{c c c ccc c ccc}
\hline\hline
Isospin & Decay channel & $\mathcal{S}(X_{1P}^{(1)})$ & $\mathcal{S}(X_{2P}^{(1)})$ & $\mathcal{S}(X_{0P}^{ (1)})$ & $\mathcal{S}(X_{0P}^{\prime(1)})$ & $\bar{\mathcal{P}}$[$\phi(2170)]$ & $\bar{\mathcal{P}}$[$\omega(2220)]$& $\bar{\mathcal{P}}$[$\phi(2350)]$ & $\bar{\mathcal{P}}$[$\phi(2470)]$ \\
\hline
$I=0$
& $\eta(548)\,f_1(1285)$                          & $\frac{-\sin{\phi_P}}{2}$  & - & - & -  & 4.6 & 4.9 & 5.0& 5.2  \\
& $\eta(548)\,f_1(1420)$                          & $\frac{\cos{\phi_P}}{2}$  & - & - & -  & 3.4 & 3.9 & 4.3& 4.5 \\
& $\eta(548)\,h_1(1170)$                          & -  & - & $\frac{-\sin{\phi_P}}{2}$  & $\frac{-\sqrt{3}\sin{\phi_P}}{2}$  & 5.4 & 5.5 & 5.6& 5.6 \\
& $\eta(548)\,h_1(1415)$                          & -  & - & $\frac{\cos{\phi_P}}{2}$  & $\frac{\sqrt{3}\cos{\phi_P}}{2}$  &  3.6 &4.1 & 4.4& 4.6\\
& $\eta(958)\,f_1(1285)$                          & $\frac{\cos{\phi_P}}{2}$   & - & - & -  & 0.0 & 1.1 & 2.6& 3.4 \\
& $\eta(958)\,f_1(1420)$                          & $\frac{\sin{\phi_P}}{2}$  & - & - &- & 0.0  & 0.0 & 0.0& 2.0\\
& $\eta(958)\,h_1(1170)$                          & -  & - & $\frac{\cos{\phi_P}}{2}$ & $\frac{\sqrt{3}\cos{\phi_P}}{2}$  & 1.7 & 3.0 & 3.6& 4.1 \\
& $\eta(958)\,h_1(1415)$                          & -  & - & $\frac{\sin{\phi_P}}{2}$ & $\frac{\sqrt{3}\sin{\phi_P}}{2}$ & 0.0 & 0.0 & 0.0& 2.3 \\
& $f_1(1420)\,\omega(782)$                        & $\frac{1}{2\sqrt{2}}$  & $\sqrt{\frac{5}{12}}$ & $\frac{1}{2}$ & $\frac{1}{2\sqrt{3}}$  & 0.0 & 1.7 & 2.8& 3.5 \\
& $f_1(1285)\,\phi(1020)$                          & $\frac{1}{2\sqrt{2}}$  & $\sqrt{\frac{5}{12}}$ & $\frac{1}{2}$ & $\frac{1}{2\sqrt{3}}$  &0.0 & 0.0 & 1.7& 2.9  \\
& $h_1(1170)\,\phi(1020)$                          & $\frac{1}{2}$  & - & - & -  & 0.0 & 2.2 & 3.1& 3.7 \\
& $h_1(1415)\,\omega(782)$                        & $\frac{1}{2}$  & - & - & -  & 0.0 & 2.0 & 3.0&3.6 \\
& $f_2(1270)\,\phi(1020)$                        & $\sqrt{\frac{5}{24}}$  & $\frac{1}{6}$ & $\sqrt{\frac{5}{12}}$ & $\frac{\sqrt{5}}{6}$  & 0.0 & 0.0 & 1.8& 3.0 \\
& $f_2(1525)\,\omega(782)$                        & $\sqrt{\frac{5}{24}}$   & $\frac{1}{6}$  & $\sqrt{\frac{5}{12}}$  & $\frac{\sqrt{5}}{6}$  &  0.0 & 0.0 & 1.7& 2.8\\
& $K\,\bar{K}_1(1270)+\mathrm{c.c.}$               & $\frac{\sin{\theta_K}}{2}$  & - & $\frac{\cos{\theta_K}}{2}$ & $\frac{\sqrt{3}\cos{\theta_K}}{2}$  & 5.1 & 5.3 & 5.4& 5.4 \\
& $K\,\bar{K}_1(1400)+\mathrm{c.c.}$               & $\frac{\cos{\theta_K}}{2}$  & - & $\frac{-\sin{\theta_K}}{2}$ & $\frac{-\sqrt{3}\sin{\theta_K}}{2}$  & 4.0 & 4.4 & 4.6&4.8 \\
& $K^\star(782)\,\bar{K}_1(1270)+\mathrm{c.c.}$                & $\frac{\sin{\theta_K}}{2\sqrt{2}}+\frac{\cos{\theta_K}}{2}$  & $\sqrt{\frac{5}{12}}\sin{\theta_K}$ & $\frac{\sin{\theta_K}}{2}$ & $\frac{\sin{\theta_K}}{2\sqrt{3}}$ &1.2 & 2.7 & 3.4& 4.0  \\
& $K^\star(782)\,\bar{K}_1(1400)+\mathrm{c.c.}$               & $\frac{\cos{\theta_K}}{2\sqrt{2}}-\frac{\sin{\theta_K}}{2}$  &$\sqrt{\frac{5}{12}}\cos{\theta_K}$ & $\frac{\cos{\theta_K}}{2}$ & $\frac{\cos{\theta_K}}{2\sqrt{3}}$ & 0.0 & 0.0 &1.8& 2.9 \\
& $K^{\star}(892)\,\bar{K}_2^{*}(1430)+\mathrm{c.c.}$   & $\sqrt{\frac{5}{24}}$  & $\frac{1}{6}$  & $\sqrt{\frac{5}{12}}$ & $\frac{\sqrt{5}}{6}$  & 0.0  & 0.0 & 1.4& 2.7\\\hline
  Isospin & Decay channel & $\mathcal{S}(X_{1P}^{(1)})$ & $\mathcal{S}(X_{2P}^{(1)})$ & $\mathcal{S}(X_{0P}^{ (1)})$ & $\mathcal{S}(X_{0P}^{\prime(1)})$ &  $\bar{\mathcal{P}}$$[\rho(2150)]$ &  $\bar{\mathcal{P}}$$[\rho(2320)]$ &  $\bar{\mathcal{P}}$$[\rho(2340)]$ &  $\bar{\mathcal{P}}$$[\rho(2490)]$ \\\hline

$I=1$
& $\eta(548)\,a_1(1260)$           &$\frac{-\sin{\phi_P}}{2}$  & - & - & -  & 5.2 & 5.3 & 5.3& 5.4 \\
& $\eta(958)\,a_1(1260)$                           & $\frac{\cos{\phi_P}}{2}$   & - & - & -  & 2.0 & 2.8 & 3.0& 3.8 \\
& $\eta(548)\,b_1(1235)$   &                        -  & - & $\frac{-\sin{\phi_P}}{2}$  & $\frac{-\sqrt{3}\sin{\phi_P}}{2}$  & 5.2 & 5.3 & 5.3& 5.4  \\
& $\eta(958)\,b_1(1235)$                          & -  & - & $\frac{\cos{\phi_P}}{2}$ & $\frac{\sqrt{3}\cos{\phi_P}}{2}$  & 2.0 & 2.8 & 3.0& 3.8 \\
& $\pi\,f_1(1420)$                          & $\frac{1}{2}$  & - & - & -  & 5.2 & 5.3 & 5.3& 5.3 \\
& $\pi\,h_1(1415)$                          & -  & - & $\frac{1}{2}$ & $\frac{\sqrt{3}}{2}$  & 5.3 & 5.3 & 5.4& 5.4 \\
& $\phi(1020)\,a_1(1260)$                          & $\frac{1}{2\sqrt{2}}$  & $\sqrt{\frac{5}{12}}$ & $\frac{1}{2}$ & $\frac{1}{2\sqrt{3}}$  & 0.0 & 2.1 & 2.3& 3.4 \\
& $\phi(1020)\,b_1(1235)$                          & $\frac{1}{2}$  & - & - & -  & 0.0 & 2.1 & 2.3& 3.4 \\
& $\rho(770)\,h_1(1415) $                     & $\frac{1}{2}$  & - & - & -  & 2.0 & 2.8 & 2.9& 3.7 \\
& $\rho(770)\,f_1(1420) $                      & $\frac{1}{2\sqrt{2}}$  & $\sqrt{\frac{5}{12}}$ & $\frac{1}{2}$ & $\frac{1}{2\sqrt{3}}$  & 1.7 & 2.6 & 2.8& 3.6 \\
& $\phi(1020)\,a_2(1320)$                         & $\sqrt{\frac{5}{24}}$  & $\frac{1}{6}$ & $\sqrt{\frac{5}{12}}$ & $\frac{\sqrt{5}}{6}$  & 0.0 & 0.0 & 0.4& 2.7  \\ 
& $\rho(770)\,f_2(1525)$                          & $\sqrt{\frac{5}{24}}$  & $\frac{1}{6}$ & $\sqrt{\frac{5}{12}}$ & $\frac{\sqrt{5}}{6}$  & 0.0 & 1.3 & 1.6& 3.0 \\ 
& $K\,\bar{K}_1(1270)+\mathrm{c.c.}$               & $\frac{\sin{\theta_K}}{2}$  & - & $\frac{\cos{\theta_K}}{2}$ & $\frac{\sqrt{3}\cos{\theta_K}}{2}$  & 5.2 & 5.3 & 5.4& 5.4 \\
& $K\,\bar{K}_1(1400)+\mathrm{c.c.}$               & $\frac{\cos{\theta_K}}{2}$  & - & $\frac{-\sin{\theta_K}}{2}$ & $\frac{-\sqrt{3}\sin{\theta_K}}{2}$  & 4.3 & 4.5 & 4.6& 4.8 \\
& $K^\star(782)\,\bar{K}_1(1270)+\mathrm{c.c.}$                & $\frac{\sin{\theta_K}}{2\sqrt{2}}+\frac{\cos{\theta_K}}{2}$  & $\sqrt{\frac{5}{12}}\sin{\theta_K}$ & $\frac{\sin{\theta_K}}{2}$ & $\frac{\sin{\theta_K}}{2\sqrt{3}}$ &2.6 & 3.3 & 3.4& 4.0  \\
& $K^\star(782)\,\bar{K}_1(1400)+\mathrm{c.c.}$               & $\frac{\cos{\theta_K}}{2\sqrt{2}}-\frac{\sin{\theta_K}}{2}$  &$\sqrt{\frac{5}{12}}\cos{\theta_K}$ & $\frac{\cos{\theta_K}}{2}$ & $\frac{\cos{\theta_K}}{2\sqrt{3}}$ & 0.0 & 1.3 & 1.7& 3.0\\
& $K^{\star}(892)\,\bar{K}_2^{*}(1430)+\mathrm{c.c.}$   & $\sqrt{\frac{5}{24}}$  & $\frac{1}{6}$  & $\sqrt{\frac{5}{12}}$ & $\frac{\sqrt{5}}{6}$  & 0.0 & 0.5 & 1.2& 2.8 \\\hline
\hline 
\end{tabular}
\end{table*}

\begin{table*}
\caption{Dominant fall-apart decay channels of the lowest $J^{PC}=2^{-+}$ tetraquark states ($I=0,1$).  All listed modes proceed in a relative $S$ wave
($\ell=0$).  ${\cal S}$ is the spin factor given in Eq.~(\ref{eq:SpinFactor}), and $\bar{\cal P}$ is $10^3$ times the phase-space factor ${\cal P}$ (in GeV$^{-1}$) given above Eq.~(\ref{eq:CMmomentum}).}
\label{tab:2mp_dominant}
\centering
\renewcommand{\arraystretch}{1.2}
\begin{tabular}{c l @{\hskip 1em} c @{\hskip 1em} c @{\hskip 1em} c c }
\hline\hline
Isospin & Decay channel & $\mathcal{S}$ ($Z_{P}^{ (2)}$) & $\mathcal{S}$ ($Z_{P}^{\prime(2)}$) & $\bar{\mathcal{P}}$$[\eta_2(2250)]$ & $\bar{\mathcal{P}}$$[\eta_2(2295)]$ \\
\hline
$I=0$
& $\eta(548)\,f_2(1270)$                          & $\frac{-\sin{\phi_P}}{2}$ &$\frac{-\sin{\phi_P}}{\sqrt{2}}$ &4.9 & 5.0 \\
& $\eta(548)\,f_2(1525)$                         & $\frac{\cos{\phi_P}}{2}$ & $\frac{\cos{\phi_P}}{\sqrt{2}}$ &3.2 & 3.4   \\
& $\eta(958)\,f_2(1270)$                         & $\frac{\cos{\phi_P}}{2}$ &$\frac{\cos{\phi_P}}{\sqrt{2}}$ &1.1& 2.0   \\
& $\eta(958)\,f_2(1525)$                        & $\frac{\sin{\phi_P}}{2}$ &$\frac{\sin{\phi_P}}{\sqrt{2}}$ & 0.0 & 0.0  \\
& $f_2(1270)\,\phi(1020)$                        & $\sqrt{\frac{3}{8}}$ &- & 0.0& 0.4  \\
& $f_2(1525)\,\omega(782)$                        & $\sqrt{\frac{3}{8}}$ &- & 0.0& 0.0  \\
& $f_1(1420)\,\omega(782)$                        & $\frac{1}{2\sqrt{2}}$ &- & 1.6& 2.3  \\
& $f_1(1285)\,\phi(1020)$                          &  $\frac{1}{2\sqrt{2}}$ &- & 0.0& 0.0  \\
& $h_1(1170)\,\phi(1020)$                          & $\frac{1}{2}$ & $\frac{1}{\sqrt{2}}$ & 2.1& 2.7  \\
& $h_1(1415)\,\omega(782)$                        & $\frac{1}{2}$ &$\frac{1}{\sqrt{2}}$ & 1.9& 2.5  \\
& $K_1(1270)\,\bar K^{*}(892)+\mathrm{c.c.}$ & $\frac{\sin{\theta_K}}{2\sqrt{2}}+\frac{\cos{\theta_K}}{2}$ & $\frac{\cos{\theta_K}}{\sqrt{2}}$ & 2.7& 3.1   \\
& $K_1(1400)\,\bar K^{*}(892)+\mathrm{c.c.}$ & $\frac{\cos{\theta_K}}{2\sqrt{2}}-\frac{\sin{\theta_K}}{2}$ & $\frac{-\sin{\theta_K}}{\sqrt{2}}$ & 0.0 & 0.5   \\
& $K\, \bar K_2^{*}(1430)+\mathrm{c.c.}$                & $\frac{1}{2}$ & $\frac{1}{\sqrt{2}}$  & 4.2 & 4.3  \\
& $K^{\star}(892)\, \bar K_2^{*}(1430)+\mathrm{c.c.}$   & $\sqrt{\frac{3}{8}}$ & - &  0.0 & 0.0 \\\hline
Isospin & Decay channel & $\mathcal{S}$ ($Z_{P}^{ (2)}$) & $\mathcal{S}$ ($Z_{P}^{\prime(2)}$) & $\bar{\mathcal{P}}$$[\pi_2(2250)]$ & $\bar{\mathcal{P}}$$[\pi_2(2290)]$ \\
\hline
$I=1$
& $\eta(548)\,a_2(1320)$                                & $\frac{-\sin{\phi_P}}{2}$ &$\frac{-\sin{\phi_P}}{\sqrt{2}}$ & 4.7 & 4.6\\ 
& $\eta(958)\,a_2(1320)$                                & $\frac{\cos{\phi_P}}{2}$  & $\frac{\cos{\phi_P}}{\sqrt{2}}$ &  0.0 & 1.0\\ 
 & $\pi\,f_2(1525)$                                &  $\frac{1}{2}$ & $\frac{1}{\sqrt{2}}$ & 4.7 & 4.8 \\ 

& $\phi(1020)\,a_1(1260)$                          & $\frac{1}{2\sqrt{2}}$ & - & 0.4 & 1.6\\ 
& $\phi(1020)\,b_1(1235)$                          & $\frac{1}{2}$ & $\frac{1}{\sqrt{2}}$ &0.4& 1.6  \\ 
& $\phi(1020)\,a_2(1320)$                          &  $\sqrt{\frac{3}{8}}$ &- & 0.0& 0.0 \\ 
& $\rho(770)\,f_2(1525)$                          & $\sqrt{\frac{3}{8}}$ &- & 0.0 & 0.0  \\ 
& $\rho(770)\,f_1(1420)$                          & $\frac{1}{2\sqrt{2}}$ &- & 1.8& 2.3\\ 
& $\rho(770)\,h_1(1415)$                          & $\frac{1}{2}$ & $\frac{1}{\sqrt{2}}$ & 2.1& 2.5 \\ 

& $K_1(1270)\,\bar K^{*}(892)+\mathrm{c.c.}$ & $\frac{\sin{\theta_K}}{2\sqrt{2}}+\frac{\cos{\theta_K}}{2}$ & $\frac{\cos{\theta_K}}{\sqrt{2}}$ & 2.7 & 3.0 \\
& $K_1(1400)\,\bar K^{*}(892)+\mathrm{c.c.}$ & $\frac{\cos{\theta_K}}{2\sqrt{2}}-\frac{\sin{\theta_K}}{2}$ & $\frac{-\sin{\theta_K}}{\sqrt{2}}$ & 0.0 & 0.0 \\
& $K\, \bar K_2^{*}(1430)+\mathrm{c.c.}$                & $\frac{1}{2}$ &$\frac{1}{\sqrt{2}}$ & 4.2 & 4.3 \\
& $K^{\star}(892)\, \bar K_2^{*}(1430)+\mathrm{c.c.}$   & $\sqrt{\frac{3}{8}}$ &- &  0.0 & 0.0\\
\hline\hline
\end{tabular}
\end{table*}

\begin{table*}
\caption{Dominant fall-apart decay channels of the lowest $J^{PC}=2^{--}$ tetraquark states ($I=0,1$).  All listed modes proceed in a relative $S$ wave
($\ell=0$).  ${\cal S}$ is the spin factor given in Eq.~(\ref{eq:SpinFactor}), and $\bar{\cal P}$ is $10^3$ times the phase-space factor ${\cal P}$ (in GeV$^{-1}$) given above Eq.~(\ref{eq:CMmomentum}).}
\label{tab:2mm_dominant}
\centering
\renewcommand{\arraystretch}{1.2}
\begin{tabular}{c l @{\hskip 1em} c @{\hskip 1em} c @{\hskip 1em} cc }
\hline\hline
Isospin & Decay channel & $\mathcal{S}$ ($X_{1P}^{(2)}$) & $\mathcal{S}$ ($X_{2P}^{(2)}$) & $\bar{\mathcal{P}}$[$\phi_2(2240))$] & $\bar{\mathcal{P}}$[$\phi_2(2440))$] \\
\hline
$I=0$
& $\eta(548)\,f_2(1270)$                          & $\frac{-\sin{\phi_P}}{2}$ & - &4.9 & 5.2  \\
& $\eta(548)\,f_2(1525)$                         & $\frac{\cos{\phi_P}}{2}$ & - &3.1 & 4.0   \\
& $\eta(958)\,f_2(1270)$                         & $\frac{\cos{\phi_P}}{2}$ &- &0.6 & 3.3   \\
& $\eta(958)\,f_2(1525)$                        & $\frac{\sin{\phi_P}}{2}$ &- & 0.0 & 0.0  \\
& $f_2(1270)\,\phi(1020)$                        & $\sqrt{\frac{3}{8}}$ &$\frac{1}{2}$ & 0.0 & 2.8  \\
& $f_2(1525)\,\omega(782)$                        & $\sqrt{\frac{3}{8}}$ &$\frac{1}{2}$ & 0.0 & 2.6 \\
& $f_1(1420)\,\omega(782)$                        & $\frac{1}{2\sqrt{2}}$ &$\frac{\sqrt{3}}{2}$ & 1.3 & 3.3 \\
& $f_1(1285)\,\phi(1020)$                          &  $\frac{1}{2\sqrt{2}}$ &$\frac{\sqrt{3}}{2}$  & 0.0 & 2.7  \\
& $h_1(1170)\,\phi(1020)$                          & $\frac{1}{2}$ & - & 1.9 & 3.6  \\
& $h_1(1415)\,\omega(782)$                        & $\frac{1}{2}$ &- & 1.7& 3.5  \\
& $K_1(1270)\,\bar K^{*}(892)+\mathrm{c.c.}$ & $\frac{\sin{\theta_K}}{2\sqrt{2}}+\frac{\cos{\theta_K}}{2}$ & $\frac{\sqrt{3}\sin{\theta_K}}{2}$ & 2.5& 3.9   \\
& $K_1(1400)\,\bar K^{*}(892)+\mathrm{c.c.}$ & $\frac{\cos{\theta_K}}{2\sqrt{2}}-\frac{\sin{\theta_K}}{2}$ & $\frac{\sqrt{3}\cos{\theta_K}}{2}$ & 0.0 & 2.7   \\
& $K\,K_2^{*}(1430)+\mathrm{c.c.}$                & $\frac{1}{2}$ & -  & 4.1 & 4.6  \\
& $K^{\star}(892)\,K_2^{*}(1430)+\mathrm{c.c.}$   & $\sqrt{\frac{3}{8}}$ & $\frac{1}{2}$ &  0.0 & 2.5 \\
\hline
Isospin & Decay channel & $\mathcal{S}$ ($X_{1P}^{ (2)}$) & $\mathcal{S}$ ($X_{2P}^{(2)}$) & $\bar{\mathcal{P}}$[$\rho_2(2250)$] & $\bar{\mathcal{P}}$[$\rho_2(2350)$] \\\hline
$I=1$
& $\eta(548)\,a_2(1320)$                                & $\frac{-\sin{\phi_P}}{2}$ &- & 4.9 & 4.6 \\ 
& $\eta(958)\,a_2(1320)$                                & $\frac{\cos{\phi_P}}{2}$  & - &  0.0 & 2.1 \\ 
 & $\pi\,f_2(1525)$                                &  $\frac{1}{2}$ & - & 4.7 & 4.9  \\ 

& $\phi(1020)\,a_1(1260)$                          & $\frac{1}{2\sqrt{2}}$ & - & 0.0 & 2.5  \\ 
& $\phi(1020)\,b_1(1235)$                          & $\frac{1}{2}$ & - &0.0& 2.5   \\ 
& $\phi(1020)\,a_2(1320)$                          &  $\sqrt{\frac{3}{8}}$ &$\frac{1}{2}$ & 0.0 & 0.9  \\ 
& $\rho(770)\,f_2(1525)$                          & $\sqrt{\frac{3}{8}}$ &$\frac{1}{2}$ & 0.0 & 1.8   \\ 
& $\rho(770)\,f_1(1420)$                          & $\frac{1}{2\sqrt{2}}$ &$\frac{\sqrt{3}}{2}$ & 1.7 & 2.9  \\ 
& $\rho(770)\,h_1(1415)$                          & $\frac{1}{2}$ & - & 2.0 & 3.0  \\ 

& $K_1(1270)\,\bar K^{*}(892)+\mathrm{c.c.}$ & $\frac{\sin{\theta_K}}{2\sqrt{2}}+\frac{\cos{\theta_K}}{2}$ & $\frac{\sqrt{3}\sin{\theta_K}}{2}$ & 2.7 & 3.4   \\
& $K_1(1400)\,\bar K^{*}(892)+\mathrm{c.c.}$ & $\frac{\cos{\theta_K}}{2\sqrt{2}}-\frac{\sin{\theta_K}}{2}$ & $\frac{\sqrt{3}\cos{\theta_K}}{2}$ & 0.0 & 1.9   \\
& $K\, \bar K_2^{*}(1430)+\mathrm{c.c.}$                & $\frac{1}{2}$ & -  & 4.2 & 4.5  \\
& $K^{\star}(892)\, \bar K_2^{*}(1430)+\mathrm{c.c.}$   & $\sqrt{\frac{3}{8}}$ & $\frac{1}{2}$ &  0.0 & 1.4 \\
\hline\hline
\end{tabular}
\end{table*}

\section{Summary and Outlook}

In this work, we have studied $P$-wave hidden-strangeness tetraquark states  $s\bar s q\bar q$ within the framework of the dynamical diquark model.  After constructing the relevant states and effective Hamiltonian, we derived the mass spectrum, highlighting the impact of spin-dependent fine-structure and isospin effects.  We obtain an excellent fit to experimentally observed resonances with the correct quantum numbers, and produce numerical results for all remaining states in the multiplet.  Most of the unconfirmed negative-parity states in the range 2.2-2.5~GeV listed by the PDG can also serve as members of this multiplet.  We encourage further detailed experimental studies of this region, particularly for exotic $J^{PC}$ values, in order to reveal this rich spectroscopy.

We also investigated strong fall-apart decay modes and identified characteristic decay patterns of these $s\bar s q\bar q$ states.  In many, but not all cases, they favor decays into kaonic final states.  Nevertheless, in most cases 2-meson final states in which one of the mesons continues to carry a hidden-strangeness component [{\it e.g.}, $f_2 (1517)$] are nearly as common, providing fertile ground for the experimental reconstruction of such states.

 States with two particular exotic quantum numbers, $1^{-+}$ and $0^{--}$, offer the most promising avenue to test the tetraquark interpretation.  The isoscalar $1^{-+}$ states predicted near 2340 and 2480~MeV decay dominantly into $K_1(1270)\bar{K}^*$ through $S$-wave fall-apart.  The $0^{--}$ case is potentially cleaner: this quantum number is absent from both conventional mesons and the hybrid spectrum, and the tetraquark multiplet predicts an isovector near 2340~MeV and an isoscalar near 2456~MeV, both with $K_1(1270)\bar{K}^*$ as the dominant $S$-wave fall-apart mode. Observation of a $0^{--}$ state near these masses would be difficult to accommodate outside of the tetraquark picture. 
 
\begin{acknowledgments}
RFL acknowledges support by the National Science Foundation (NSF) under Grant No.\ PHY-2405262.  SJ acknowledges support by the U.S.\ Department of Energy ExoHad Topical Collaboration under Grant No.\ DE-SC0023598.  This work contributes to the goals of the ExoHad Collaboration.
\end{acknowledgments}

\bibliographystyle{apsrev4-2}
\bibliography{main}
 
\end{document}